\documentclass[12pt,dvipsnames]{article}
\usepackage[top=80pt,bottom=85pt,left=75pt,right=75pt]{geometry}

\pdfoutput=1
\usepackage[utf8]{inputenc}
\usepackage[T1]{fontenc} 
\usepackage[english]{babel} 
\usepackage{cancel}
\usepackage{braket}
\usepackage{csquotes}
\usepackage[dvipsnames]{xcolor} 
\usepackage{physics}
\usepackage{tabularx} 
\usepackage{booktabs} 
\usepackage{dsfont}
\usepackage{rotating}
\usepackage{multirow} 
\usepackage{comment} 
\usepackage{amsmath}
\usepackage{bbm}
\usepackage{todonotes}
\usepackage{tikz}
\usetikzlibrary{math} 
\usetikzlibrary{3d} 
\usepackage{tikz-3dplot}
\usepackage{tikz-cd} 
\usepackage{microtype}
\usepackage{amsthm}
\usepackage{dsfont}
\usepackage{enumitem}
\usepackage[strict]{changepage}

\usepackage{tikz}
\usepackage{asymptote}
\usetikzlibrary{3d}
\usetikzlibrary{calc}
\usepackage{pgfplots}
\usepackage{tikz-3dplot}

\pgfplotsset{
	compat=newest
}

\tikzcdset{
  cells={font=\everymath\expandafter{\the\everymath\displaystyle}},
}

\usepackage{amssymb} 
\usepackage{subcaption} 
\DeclareCaptionFormat{custom}

\usepackage[pdftex,colorlinks=true]{hyperref} 
\hypersetup{urlcolor=MidnightBlue, citecolor=red, linkcolor=MidnightBlue}

\usepackage{float} 
\usepackage[capitalise]{cleveref}

\newcolumntype{E}{>{\hfil$}p{0.65cm}<{$\hfil}}

\newcolumntype{D}{>{\hfil$}p{7.4cm}<{$\hfil}}
\newcolumntype{C}{>{\hfil$}p{3cm}<{$\hfil}}
\newcolumntype{P}{>{\hfil$}p{7.7cm}<{$\hfil}}
\newcolumntype{F}{>{\hfil$}p{5.7cm}<{$\hfil}}
\newcolumntype{L}{>{\hfil$}p{2.6cm}<{$\hfil}}
\newcolumntype{S}{>{\hfil$}p{1.8cm}<{$\hfil}}
\newcolumntype{R}{>{\hfil$}p{5.2cm}<{$\hfil}}
\newcolumntype{U}{>{\hfil$}p{4.2cm}<{$\hfil}}
\newcolumntype{Q}{>{\hfil$}p{6.4cm}<{$\hfil}}
\newcolumntype{T}{>{\hfil$}p{1.9cm}<{$\hfil}}
\newcolumntype{V}{>{\hfil$}p{5.8cm}<{$\hfil}}
\newcolumntype{H}{>{\hfil$}p{1.8cm}<{$\hfil}}
\newcolumntype{A}{>{\hfil$}p{6cm}<{$\hfil}}
\newcolumntype{B}{>{\hfil$}p{2cm}<{$\hfil}}

\tikzset{7brane/.style={circle, draw=black, fill=black,ultra thick,inner sep=1.5 pt, minimum size=1 pt,}, c/.default={4pt}}

\makeatletter
\newcommand\xleftrightarrow[2][]{%
  \ext@arrow 9999{\longleftrightarrowfill@}{#1}{#2}}
\newcommand\longleftrightarrowfill@{%
  \arrowfill@\leftarrow\relbar\rightarrow}
\makeatother

\renewcommand{\vec}[1]{\mathbf{#1}}

\newcommand{\C}{\mathbb{C}}

\definecolor{bittersweet}{rgb}{1.0, 0.44, 0.37}

\numberwithin{equation}{section}

\usepackage[maxnames=99,citestyle=numeric-comp,sorting=none,doi=true]{biblatex}
\allowdisplaybreaks[1] 

\hypersetup{
pdftitle={The Geometry of GTPs and 5d SCFTs},    
pdfauthor={\textcopyright\ Guillermo Arias Tamargo, Diego Rodr\'iguez G\'omez, Sebasti\'an Franco},     
pdfsubject={HEP},   
pdfcreator={pdfLaTex},   
pdfproducer={LaTex}, 
pdfkeywords={},
colorlinks=true,
linkcolor=blue,
urlcolor=blue,
filecolor=blue,
citecolor=blue,
linktocpage=true
}

\definecolor{cambridgeblue}{rgb}{0.64, 0.76, 0.68}
\definecolor{caribbeangreen}{rgb}{0.0, 0.8, 0.6}
\definecolor{celadon}{rgb}{0.67, 0.88, 0.69}
\definecolor{champagne}{rgb}{0.97, 0.91, 0.81}
\definecolor{cream}{rgb}{1.0, 0.99, 0.82}
\definecolor{cyan(process)}{rgb}{0.0, 0.72, 0.92}
\definecolor{brilliantlavender}{rgb}{0.96, 0.73, 1.0}
\definecolor{candypink}{rgb}{0.89, 0.44, 0.48}

\begin{document}

\begin{titlepage}

\vspace{-3cm}

\centerline{\LARGE \bf The Geometry of GTPs and 5d SCFTs}
\medskip

\vspace{.9truecm}

\centerline{
    { Guillermo Arias-Tamargo}\textsuperscript{\it a} \footnote{guillermo.arias.tam@gmail.com},
    {Sebasti\'an Franco}\textsuperscript{\it b,c,d} \footnote{sfranco@ccny.cuny.edu} and  {Diego Rodr\'iguez-G\'omez}\textsuperscript{\it e,f} \footnote{d.rodriguez.gomez@uniovi.es}}

\vspace{0.8cm}

\begin{small}
\centerline{\textsuperscript{\it a}\it  Theoretical Physics Group, The Blackett Laboratory, Imperial College London,}

\centerline{\it  Prince Consort Road London, SW7 2AZ, UK}
\medskip

\centerline{\textsuperscript{\it b}\it  Physics Department, The City College of the CUNY,}

\centerline{\it  160 Convent Avenue, New York, NY 10031, USA}

\medskip

\centerline{\textsuperscript{\it c}\it  Physics Program and \textsuperscript{\it d}\it Initiative for the Theoretical Sciences,}

\centerline{\it  The Graduate School and University Center, The City University of New York}

\centerline{\it  365 Fifth Avenue, New York NY 10016, USA}

\medskip

\centerline{\textsuperscript{\it e}\it  Department of Physics, Universidad de Oviedo,}

\centerline{\it  C/ Federico Garcia Lorca 18, 33007 Oviedo, Spain}

\medskip

\centerline{\textsuperscript{\it f}\it  Instituto Universitario de Ciencias y Tecnologias Espaciales de Asturias (ICTEA),}

\centerline{\it  C/ de la Independencia 13, 33004 Oviedo, Spain}
\end{small}

\vspace{0.5cm}

\centerline{\small{\bf Abstract} }

\vspace{0.2cm}

\begin{center}
\begin{minipage}[h]{0.93\textwidth}
\linespread{1.0}\selectfont
We make progress in understanding the geometry associated to the Generalized Toric Polygons (GTPs) encoding the Physics of 5d Superconformal Field Theories (SCFTs), by exploiting the connection between Hanany-Witten transitions and the mathematical notion of polytope mutations. From this correspondence, it follows that the singular geometry associated to a GTP is identical to that obtained by regarding it as a standard toric diagram, but with some of its resolutions frozen in way that can be determined from the invariance of the so-called period under mutations. We propose the invariance of the period as a new criterion for distinguishing inequivalent brane webs, which allows us to resolve a puzzle posed in the literature. A second mutation invariant is the Hilbert Series of the geometry. We employ this invariant to perform quantitative checks of our ideas by computing the Hilbert Series of the BPS quivers associated to theories related by mutation. Lastly, we discuss the physical interpretation of a mathematical result ensuring the existence of a flat fibration over $\mathbb{P}^1$ interpolating between geometries connected by mutation, which we identify with recently introduced deformations of the corresponding BPS quivers.

\end{minipage}

\end{center}
\end{titlepage}
\newpage

\tableofcontents 

\setcounter{footnote}{0} 
\setcounter{page}{1}

\section{Introduction}

Constructing interacting UV complete Quantum Field Theories (QFTs) in $d>4$ is notoriously hard. In the particular case of $d=5$ it is fair to say that the existence of consistent non-supersymmetric interacting QFTs has not been clearly established (see \textit{e.g.} \cite{BenettiGenolini:2019zth,Bertolini:2021cew, DeCesare:2021pfb,Bertolini:2022osy,DeCesare:2022obt, Akhond:2023vlb}). The situation is much better for the case of supersymmetric theories, since in that case, using the power of String/M theory, it is possible to conclusively construct 5d interacting and consistent SCFTs. It turns out that these theories are very interesting. Since 5d SCFTs do not have marginal deformations, they are intrinsically strongly coupled and often exhibit rather exotic properties, including for instance global symmetries of exceptional type. Moreover, upon compactification, they can provide new perspectives on strong coupling phenomena in lower dimensions.

Within String/M theory there are various approaches to constructing 5d SCFTs, perhaps most saliently through geometric engineering them in M-theory on a 3d Calabi-Yau (CY) $\mathcal{M}$ \cite{Morrison:1996xf,Douglas:1996xp,Intriligator:1997pq} (see e.g. \cite{DelZotto:2017pti,Xie:2017pfl,Alexandrov:2017mgi,Jefferson:2017ahm,Jefferson:2018irk,Bhardwaj:2018yhy,Bhardwaj:2018vuu,Apruzzi:2018nre,Closset:2018bjz,Bhardwaj:2019jtr,Apruzzi:2019vpe,Apruzzi:2019opn,Apruzzi:2019enx,Bhardwaj:2019xeg,Saxena:2020ltf,Apruzzi:2019kgb,DeMarco:2023irn,Dimofte:2011ju} for recent work) and on the worldvolume of 5-brane webs in Type IIB String Theory \cite{Aharony:1997bh} (see also \cite{Benini:2009gi,Bergman:2012kr,Bergman:2013aca,DeWolfe:1998eu,Feng:2005gw,DeWolfe:1999hj}). In recent years, significant efforts have been devoted to understanding 5d QFTs using these methods, as well analytic tools such as supersymmetric localization. 

Regarding the 5-brane web avatar, it has been realized that, in order to explicitly show properties of 5d SCFTs such as the full Higgs branch, it is necessary to think of the web as ending on 7-branes. This has two immediate consequences. On one hand, when more than one 5-brane ends on a 7-brane, the 7-brane imposes non-trivial supersymmetric boundary conditions --which go by the name of \textit{s-rule}-- which constrain the Coulomb branch of the 5d SCFTs. On the other hand, since the position of the 7-brane along the prong of the web defined by the 5-branes ending on it (the leg of the web) is not a parameter in the 5d SCFT, one may imagine crossing the 7-brane to the other side of the web. Since 7-branes come with a branch cut, as it sweeps part of the web it changes the type of a subset of the remaining 7-branes and 5-branes, sometimes leading to 5-brane creation through the celebrated Hanany-Witten (HW) effect \cite{Hanany:1996ie}. Thus, one may have two different looking webs describing the same 5d SCFT. Indeed, due to this fact, the program of classifying 5d SCFTs through classifying the possible brane webs has proven to be very hard.

When all of external legs of the web consist of a single 5-brane ending on the corresponding 7-brane, the latter can actually be neglected. In that case, it turns out that there is a relation with the geometric engineering approach when $\mathcal{M}$ is toric. Reducing on a $T^2$ inside the toric fiber, M-theory becomes Type IIB String Theory with 5-branes along the locus where the $T^2$ pinches off, which is precisely the corresponding 5-brane web. Thus, the brane web is the toric skeleton of the CY$_3$, and can be regarded as the diagram graph-dual to the toric diagram of $\mathcal{M}$. However, in the generic case of an arbitrary number of 5-branes ending on each 7-brane, the geometric engineering description is not known. The notion of {\it Generalized Toric Polygon} (GTP) has been developed to describe these more general cases \cite{Benini:2009gi,vanBeest:2020kou,VanBeest:2020kxw}. GTPs look like standard toric diagrams but are decorated with white and black dots, encoding how 5-branes may terminate on the same 7-brane if the corresponding external segments in the GTP are separated by a white dot. Moreover, a rule for supersymmetrically triangulate (actually ``polygonate'') the interior of a GTP according to the $s$-rule has been proposed. While GTPs look very close to standard toric diagrams (and reduce to these when all dots are black), their geometric interpretation has not been fully established.\footnote{Recent progress, on which we will elaborate, includes \cite{Bourget:2023wlb,Franco:2023flw,Franco:2023mkw,Cremonesi:2023psg}.} The purpose of this paper is to make progress on these questions, notably by establishing a connection to a branch of Mathematics that has emerged over the last decade, largely in parallel with the interest in the Physics community regarding 5d SCFTs, and deals with essentially the same problem.

Specifically, as suggested in \cite{Franco:2023flw}, in this paper we will argue that the mathematical notion of mutation turns out to precisely correspond to HW crossing of branes. Note that HW transitions typically lead to GTPs, even if the starting point is a standard toric diagram. Yet, from the mathematical point of view, the decoration of the GTP with white dots is irrelevant: it simply describes the singular toric variety as if no decoration was present. Moreover, two toric varieties whose toric polytopes are related by mutation can be regarded as members of a flat fibration over $\mathbb{P}^1$, which implies that, geometrically, the HW transition can be regarded as a deformation of the starting CY$_3$. This establishes that the (singular) toric variety geometrically engineering the 5d SCFT in M-theory is simply that associated to the GTP forgetting its white dot decoration. Exploiting the established invariants under mutation developed in the mathematical literature --which will also enable us to refine previous classifications of 5-brane webs-- we will demonstrate that the white dot decoration has the effect of freezing possible resolutions of the geometry in a precise way. 

Moreover, we will offer quantitative evidence of this proposal by studying the BPS quiver of the 5d SCFT compactified on $S^1$. This quiver is identical to the fractional brane quiver for Type IIB D3-branes probing the toric CY$_3$ variety $\mathcal{M}$. Thus, following the proposal above, for an arbitrary 5d theory encoded in an arbitrary GTP, we will read off the BPS quiver by simply forgetting the decoration of the GTP and regarding it as a standard toric diagram. This is an easy task, given the substantial brane tiling machinery developed for this purpose \cite{Franco:2005rj,Franco:2005sm}. We will see that the partition function counting gauge-invariant operators of the BPS quiver matches, upon using the appropriate prescription, the Hilbert series of the variety. The latter has been shown to be invariant under mutation, and the agreement of quiver partition functions will be a non-trivial check of our ideas. 

This paper is organized as follows. In Section \ref{sec:geoingandwebs} we give a lightning review of the two approaches to construct 5d SCFTs which we consider, namely geometrically engineering them in M-theory and on 5-brane webs in Type IIB String Theory. In Section \ref{sec:HWandMutation}, we describe the identification of Hanany-Witten transitions in 5-brane webs with the mathematical notion of polytope mutation. This allows us to introduce two quantities, the so-called period and the Hilbert series, which are invariant under this transformation. In Section \ref{sec:periods}, we explore the physical implications of the invariance of the period. This will allow us to recover, from first principles, the recipe for the Seiberg-Witten (SW) curves in \cite{Kim:2014nqa}, as well as to resolve a puzzle posed in \cite{Arias-Tamargo:2022qgb} concerning the classification of 5-brane webs. In Section \ref{sec:quivers}, we turn to the invariance of the Hilbert series, which provides a quantitative consistency check of the perspective advocated in this paper. We offer further examples in Section \ref{sec:examples}. We conclude in Section \ref{sec:discussion} with a summary of our results and a discussion of the open problems and future perspectives raised by our work.

\section{Geometric engineering and brane webs}\label{sec:geoingandwebs}

In this section we review three of the main approaches to construct 5d $\mathcal{N}=1$ SCFTs and their interrelation. First we begin by considering their geometric engineering via M-theory on a local Calabi-Yau threefold. Second, we study their construction via webs of 5-branes in flat space; in simple cases the correspondence with the geometric setup is well known. Last, we will review the mirror construction of the original M-theory CY.

\subsection{M-theory on CY$_3$} \label{sec:M_theory_geometry}

Let us now focus on the first of these three setups, namely consider M-theory on a three (complex) dimensional, non-compact, canonical, Gorenstein singularity. Following the standard terminology in the literature, we will refer to it as a local CY$_3$, and denote it by $\mathcal{M}$. Since the geometry is non-compact, gravity is decoupled, and this system is described by a 5d SCFT living in the 5 directions transverse to the singularity \cite{Morrison:1996xf,Douglas:1996xp,Intriligator:1997pq}.

In this situation, there is a correspondence between the field theory and the geometrical data. Let us denote by $\widetilde{\mathcal{M}}$ a (partial) crepant resolution of $\mathcal{M}$. This geometry has a set of non-compact divisors that arise directly from those in $\mathcal{M}$, as well as a set of compact exceptional divisors that arise from the resolution. Together, they form the extended K\"ahler cone of $\widetilde{\mathcal{M}}$, which we denote $\mathcal{K}(\widetilde{\mathcal{M}})$. In the low energy field theory description, they correspond to the 5d extended Coulomb branch; more precisely compact cycles correspond to proper Coulomb branch VEVs, while non-compact cycles map to (real) mass deformations. In fact, real masses can be understood as VEVs for scalars in background vector multiplets for global symmetries, hence the name of extended Coulomb branch.

In the rest of the paper, we will be focusing on the case where the Calabi-Yau is toric, which translates to the field theory having at least a $\mathrm{U}(1)^3$ global symmetry. This means that generically complicated geometric notions can be simplified in terms of combinatorics, since the complexified torus $(\mathbb{C}^*)^3$ acts on a dense subvariety of $\mathcal{M}$. Indeed, among many other features, this allows to trade the defining equations of $\mathcal{M}$ as an algebraic variety for its toric fan, which is a collection of vectors in a $\mathbb{Z}^3$ lattice specifying the weights of the torus action. In these terms, the Calabi-Yau condition is seen as the fact that all these vectors end on the same 2d plane. This further streamlines the combinatorics, as it is now sufficient to focus on the 2d polygon defined by the endpoints of the vectors in the toric fan.

In these terms, the discussion regarding the extended Coulomb branch of the 5d theory simplifies. The singular Calabi-Yau $\mathcal{M}$ corresponds to a polygon such that only external lattice points (that is, along the perimeter) are joined by edges. Its resolution $\widetilde{\mathcal{M}}$ corresponds to a triangulation of the polytope for $\mathcal{M}$; the variety is completely resolved if every internal point is joined by lines of the polytope. The lattice points themselves correspond to divisors of the geometry.\footnote{One should note that when modding out by principal divisors in order to find the divisor class group, not all points correspond to independent divisor classes.} From the discussion above, it follows that internal points are mapped to proper Coulomb branch deformations of the 5d theory (compact cycles), while external points correspond to mass deformations (non-compact cycles). The number of internal points of the polygon is the rank of the 5d SCFT. 

BPS states of the 5d theory can be understood in M-theory. M2-branes wrapping certain holomorphic 2-cycles are identified with W-bosons and instanton particles, and M5-branes wrapping 4-cycles give rise to instanton strings. Making this map more precise would require introducing a ruling for the compact divisors, but we will not do that here as we will not make use of it. Instead, it is easier to look at the BPS states from the dual Type IIA picture as follows.

We can consider compactifying one of the field theory directions on a circle, obtaining a so-called 4d KK theory, described by Type IIA String Theory on the CY$_3$. The different BPS states of the theory correspond to the bound states of D0-D2-D4 branes supersymmetrically wrapped in $\widetilde{\mathcal{M}}$. Such bound states are captured by the {\it BPS quiver}, which coincides with the fractional brane quiver of the CY$_3$ singularity \cite{closset20205d}. This object has been thoroughly studied in the past in the case of toric CY$_3$'s. 
This allows us to import the heavy machinery developed to describe Type IIB D3-branes probing toric CY$_3$'s in terms of brane tilings \cite{Franco:2005sm} (see \cite{Kennaway:2007tq,Yamazaki:2008bt} for comprehensive reviews) to construct the BPS quiver for (compactified) 5d SCFTs geometrically engineered by the CY$_3$.

One interesting consequence of this correspondence is that it is possible to go both ways. We just discussed how to reach the BPS quiver starting from the geometry. But taking the BPS quiver as a starting point, one can also obtain the CY$_3$ geometry $\mathcal{M}$ that engineers the 5d SCFT in M-theory. Indeed, given a BPS quiver we may regard it as a 4d $\mathcal{N}=1$ quiver for D3-branes probing a CY$_3$, and computing the geometry of the moduli space of vacua of the 4d theory, we recover the original Calabi-Yau.

\subsubsection{The $E_1$ example} 

\begin{figure}[h]
	\centering
	\includegraphics[height=3cm]{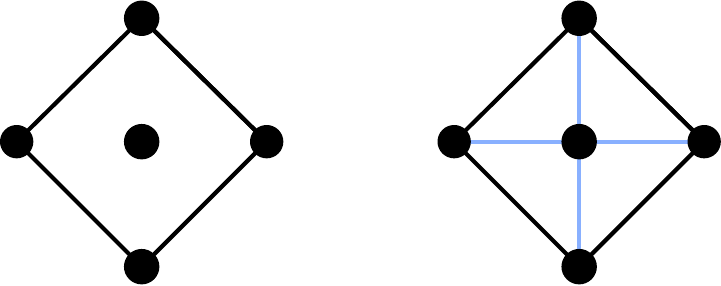}
\caption{Toric diagram for the $E_1$ theory before and after the resolution.}
	\label{toric_E1}
\end{figure}

Let us illustrate these concepts with some detail in the simple example of the $E_1$ theory, to which we will come back throughout this work. This theory is engineered by M-theory on the complex cone over the Hirzebruch surface $\mathbb{F}_0=\mathbb{P}^1\times\mathbb{P}^1$, and will be denoted by $C(\mathbb{F}_0)$. Both $\mathbb{F}_0$ and $C(\mathbb{F}_0)$ are toric varieties.\footnote{In the Physics literature CY$_3$ geometries which are a complex cone over a 2d base $C(B)$ are often denoted by simply the 2d base $B$.} The (dual of the) toric fan of $C(\mathbb{F}_0)$ can be projected to the hyperplane at height 1 obtaining a 2d integral polygon known as the toric diagram, which we depict in Figure \ref{toric_E1}.

This specifies for us the weights of the toric action, which is then $(u,v,w,t) \to (t_1 t_3 u, t_2 t_3 v, \frac{t_3}{t_1} w , \frac{t_3}{t_2} t)$. From this we recognize the equations corresponding to two copies of $\mathbb{P}^1$ over the complex plane. This variety can also be regarded as a $\mathbb{Z}_2$ orbifold of the conifold.

This singular variety has four toric divisors $D_i, i=1,\dots,4$ given by setting each of the 4 complex coordinates equal to zero. After the blowup, we have an additional exceptional divisor $D_e$. These divisors are not all linearly independent: after modding out by principal divisors, one finds the relations
\begin{align}
    [D_1] - [D_3] = 0\,,\nonumber\\
    [D_2] - \left[ D_4 \right] = 0\,,\\
    [D_e] + [D_1] + [D_2] + [D_3] + [D_4] = 0\,.\nonumber
\end{align}
Therefore we conclude that there are two linearly independent divisors, which we take to be $[D_e]$ and, say, $[D_1]$. One can compute their volumes by integrating the K\"ahler form of $C(\mathbb{F}_0)$. The volume of the (compact) exceptional divisor corresponds to the proper Coulomb branch modulus, and the volume of the (non-compact) toric divisor to the extended Coulomb branch modulus; together, they generate the extended K\"ahler cone of $C(\mathbb{F}_0)$.\footnote{More precisely the K\"ahler parameters live in $H^2(C(\mathbb{F}_0),\mathbb{R})$. The divisors live in $H_4(C(\mathbb{F}_0),\mathbb{Z})$ and their Poincar\'e duals in $H^2(C(\mathbb{F}_0),\mathbb{Z})$, and taking into account the volume leads to the desired coefficients for the cohomology group.} Indeed, we find that its dimension is 2, as expected for a theory with gauge and global symmetry of rank 1.

\begin{figure}[h]
	\centering
	\includegraphics[height=5cm]{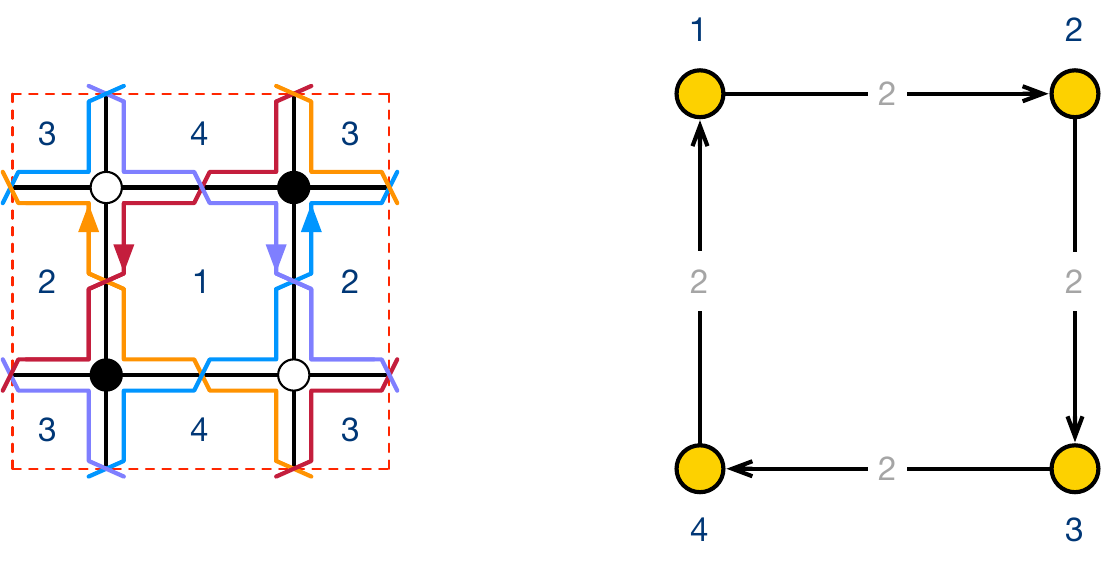}
\caption{Brane tiling and quiver diagram for $C(\mathbb{F}_0)$.}
	\label{tiling_quiver_F0}
\end{figure}

The BPS quiver coincides with the fractional brane quiver for D3-branes probing $C(\mathbb{F}_0)$. This quiver and its superpotential can be easily computed using the standard technology of dimer models, as shown in Figure \ref{tiling_quiver_F0}.

The corresponding superpotential is
\begin{equation}
W=-X_{12}X_{23}X_{34}X_{41}+X_{41}Y_{12}X_{23}Y_{34}+X_{12}Y_{23}X_{34}Y_{41}-Y_{12}Y_{23}Y_{34}Y_{41}\,.
\label{W_F0_1}
\end{equation}

From the field content and the superpotential, one can compute the complete F-terms satisfied by the chiral fields, and it is possible to see that the moduli space of the quiver in \ref{tiling_quiver_F0} precisely reproduces the toric diagram in \ref{toric_E1}.

It is worth noting that the fractional brane quiver associated to a toric CY cone is not unique, as there can be multiple Seiberg dual phases. In the example of $C(\mathbb{F}_0)$ at hand there is another phase, obtaining by Seiberg-dualizing any node in the quiver in \ref{tiling_quiver_F0} (all nodes are equivalent).

\subsection{Brane webs and GTPs} \label{sec:Brane_webs_GTPs}

The duality between M-theory on $T^2$ and Type IIB on $S^1$ implies that a 5d theory can be engineered both via pure geometry in M-theory, as described above, or through a brane web in flat space in Type IIB \cite{Leung:1997tw}. Translating between the two setups is a well understood problem when the geometry is a toric CY$_3$. In this case, the edges of the toric diagram, which correspond to 2d faces of the toric fan, signal that two of the three circles in $(\mathbb{C}^*)^3$ are pinched. In the dual Type IIB setup, this pinching is seen as a discontinuity of the $B_2$ and $C_2$ fields which, in turn, signals the presence of a charged 5-brane. 

\begin{figure}[h]
	\centering
	\includegraphics[height=4cm]{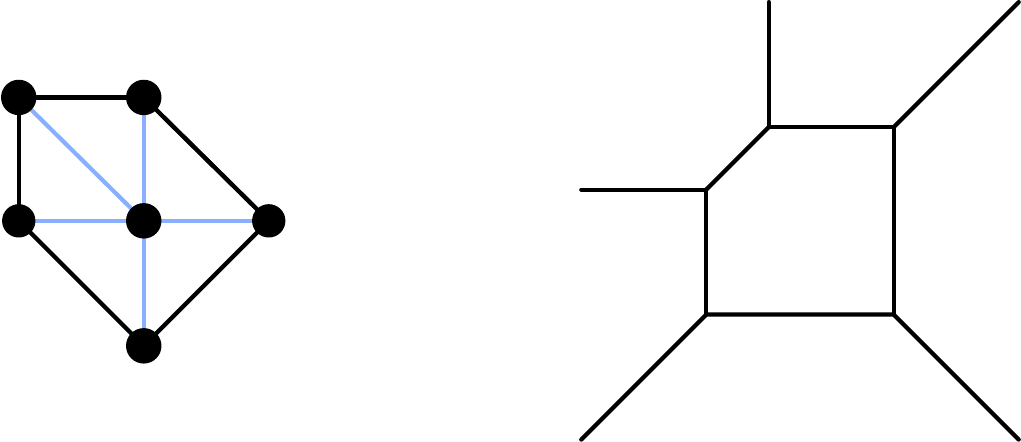}
\caption{Toric diagram and brane web for $SU(2)$ SQCD with 1 flavor.}
	\label{web_toric_SU2 - 1 flavor}
\end{figure}

More precisely, if a $(p,q)$ complex 2-cycle of the torus pinches, it will correspond to a $(p,q)$-5-brane. The conclusion is that, given a toric diagram for the M-theory CY$_3$, the dual of this graph (sometimes referred to as the toric skeleton) is directly the 5-brane web in Type IIB. As an illustration, Figure \ref{web_toric_SU2 - 1 flavor} shows the toric diagram and brane web for $SU(2)$ SQCD with 1 flavor.

More generally, 5-brane webs in Type IIB String Theory have been thoroughly studied by themselves starting with \cite{Aharony:1997bh}. Choosing the axio-dilaton $\tau=i$, a $(p,q)$ 5-brane wraps $(01234)$ and looks like a segment or line with slope $q/p$ on the $(56)$ plane. 5-branes can join, provided $(p,q)$ charge is conserved at every intersection, forming a web on the $(56)$ plane. Note that if $(p,q)$ are not coprime, the corresponding segment describes $n=\rm{gcd}(p,q)$ 5-branes with charges $(p/n,q/n)$. To fix some notation, we will always consider the 5-brane charges $(p,q)$ to be coprime, and we will refer to the number of parallel 5-branes $n$ as the \emph{multiplicity} of the leg.

These configurations can be extended by adding 7-branes spanning $(01234789)$ and located at a point on the plane of the web with no further breaking of supersymmetry, such that every leg of the web can terminate on an appropriate 7-brane. To be explicit, a 7-brane is described by its $[p,q]$ charge. Once again we assume ${\rm gcd}(p,q)=1$ so that we describe a single 7-brane, on which several $(p,q)$ 5-branes can end; and we write the total charge vector of the leg as $n(p,q)$. The addition of the 7-branes allows the visualization of the Higgs branch of the 5d theory as ``sliding'' sub-webs in the $(789)$ directions. Moreover, the 7-branes impose boundary conditions on the web which restrict the number of Coulomb branch deformations and are constrained by the supersymmetry of the configuration through the so-called $s$-rule. To see this, let us consider a 5-brane web with $L$ external legs, and where each external leg ends on the corresponding $[p_i,q_i]$ 7-brane and has multiplicity $n_i$, with $i=1,\cdots, L$. Thus, we can denote each 7-brane by a vector $\ell_i=[p_i,q_i]$, and for definiteness, choose to label them anti-clock wise. Note that in these conventions, charge conservation reduces to $\sum n_i\ell_i=0$. The self-intersection of the web is defined as \cite{Iqbal:1998xb,Bergman:2020myx}\footnote{It is important to stress that the formula in \eqref{eq:I} is valid for irreducible webs, that is, those which are not the superimposition of various individual webs.}
\begin{equation}
\label{eq:I}
    \mathcal{I}=\left|\sum_{i\leq j} n_i\,n_j\,\langle\ell_i|\ell_j\rangle\right|-\sum_{i}n_i^2\,,\qquad \langle \ell_i|\ell_j\rangle={\rm det}(\vec{\ell}_i,\vec{\ell}_j)\,.
\end{equation}
The condition for the web to be supersymmetric is
\begin{align}
    \mathcal{I}\ge-2\,.
\end{align}
There is a relation between the self-intersection of the web and the dimension of the Coulomb branch of the 5d theory, given by 
\begin{align}
\label{eq:dC}
    d_C=\frac{\mathcal{I}+2}{2}\,,
\end{align}
so that the so that the SUSY condition translates to the Coulomb branch having non-negative dimension $d_C\geq 0$.  

Since 7-branes are point like in the $(56)$ plane, they come with a branch cut for the axio-dilaton that they source, which is specified by their $[p,q]$ charge. The associated monodromy matrix is\footnote{We use conventions such that when the branch cut sweeps counter-clockwise an $[r,s]$ 7-brane, the latter gets transformed into an $M_{(p,q)}[r,s]^T$ 7-brane. If, instead, the cut sweeps the 7-brane clockwise, it acts with $M_{(p,q)}^{-1}$.}
\begin{equation}
    M_{(p,q)}=\left(\begin{array}{cc}1-pq & p^2 \\ -q^2 & 1+pq\end{array}\right)\,.
\end{equation}

We assume a ``standard presentation'' for the web, where all branch cuts are assumed to run 
away from the web, not crossing any brane. For instance, we can take them to run ``radially'' along the direction of the prong corresponding to the leg. 

Given a web ending on 7-branes, there are two possible types of motions of the 7-branes: a) we can move the 7-branes changing the asymptotic positions of the external legs attached to them, or b) we can move the 7-branes along the external legs without changing their asymptotic position. Let us discuss the consequences of each of these alternatives.

\begin{enumerate}[label=(\alph*)]
    \item Moving the 7-branes changing the asymptotic position of the external legs corresponds to a mass deformation of the 5d SCFT. Interestingly, these deformations may open Higgs branch directions. Indeed, moving two 7-branes such that two or more external legs of the brane web coincide leads to a 5-brane segment which is free to move in the direction perpendicular to the plane of the web, corresponding to a Higgs branch VEV in the 5d theory, as discussed above. Integrating out this massive mode we are left with a web where two parallel 5-branes end in the same 7-brane. This results on some moduli of the theory becoming frozen: in fact, together with the requirement that every time that 5-branes meet the junction needs to be supersymmetric, more moduli can get frozen beyond the position of the external legs. As an example, consider the webs shown in Figure \ref{web_1}. Requiring that the rightmost junction of the web on the RHS satisfies $\mathcal{I}\ge-2$ implies that there is only one free Coulomb branch moduli.

\begin{figure}[h]
	\centering
	\includegraphics[height=4cm]{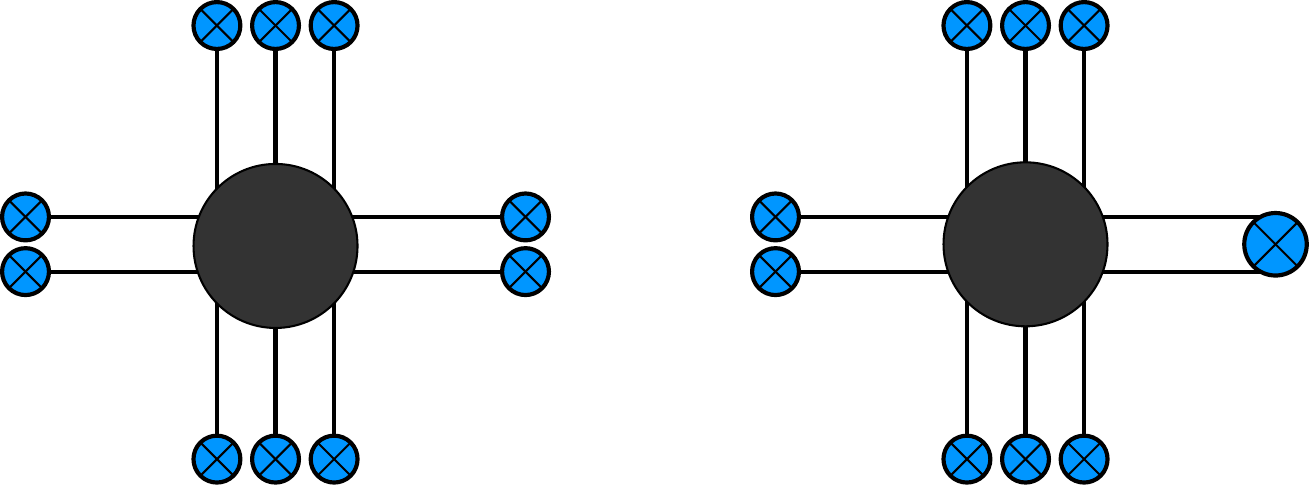}
\caption{Brane web where two 5-branes end on the same 7-brane after a Higgs branch flow.}
	\label{web_1}
\end{figure}
    
    \item Consider instead moving a 7-brane along the corresponding external leg. Such motion does not correspond to an observable of the 5d theory. In particular, we can consider moving the 7-brane across the web. In order to keep its monodromy cut pointing in outwards, we have to rotate it into the opposite direction, sweeping in the process ``half of the web'' and transforming the 7-branes that are crossed accordingly. To be more explicit, suppose we cross the $i$-th brane with charge vector $\ell_i$. Then, for the transformed 7-branes, their charge vector transforms as $\ell_j\rightarrow M_{\ell_i}\ell_j=\ell_j+\langle \ell_i|\ell_j\rangle\,\ell_j$, so that the full transformation of the web is

    \begin{equation}
        \label{HWtransf}
        \ell_j\rightarrow \begin{cases}\ell_j\,,\qquad j\leq i\,, \\  -\ell_i\,, \qquad j=i\,,\\ \ell_j+\langle \ell_i|\ell_i\rangle\,\ell_j\,,\qquad j>i\,.\end{cases}
    \end{equation}
    
    As a result, through the Hanany-Witten effect, the number of 5-branes ending on the 7-brane needs to be changed so as to satisfy charge conservation. More explicitly, the multiplicities must transform as

     \begin{equation}
        \label{HWtransfmult}
        n_j\rightarrow \begin{cases}n_j\,,\qquad j\leq i\,, \\ -n_i+\sum_{k>j} n_k \,\langle \ell_i|\ell_k\rangle \,,\qquad j=i\,, \\ n_j\,,\qquad j>i\,.\end{cases}
    \end{equation}
    
    Figure \ref{web_2} shows an example of this process, which starts from the web for the $T_3$ theory and moves one of the [1,0] 7-branes from the left to the right of the web.

\begin{figure}[h]
	\centering
	\includegraphics[height=4cm]{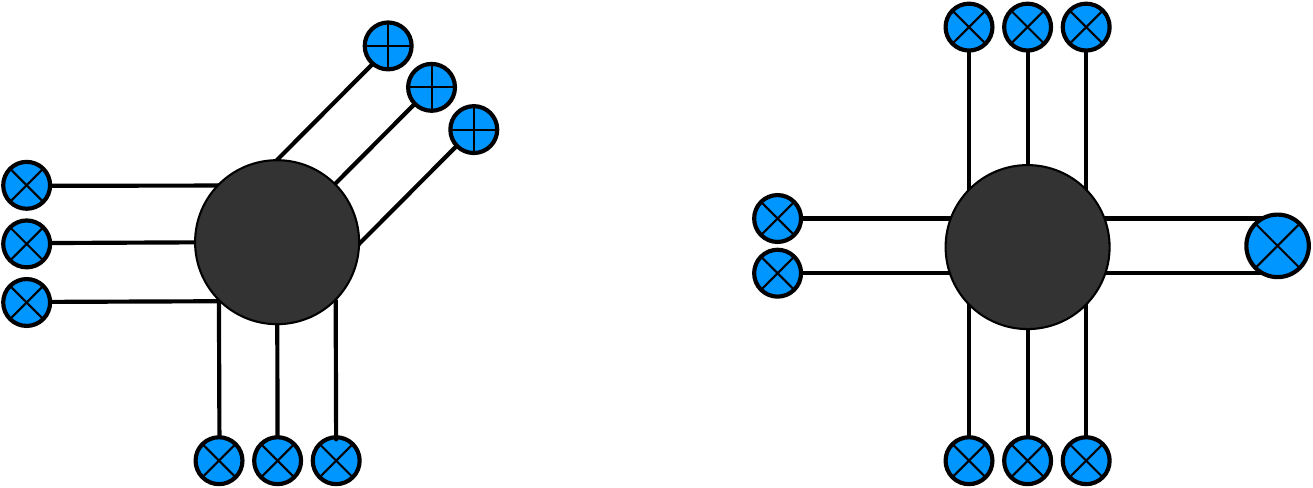}
\caption{Brane web where two 5-branes end on the same 7-brane after a HW transition.}
	\label{web_2}
\end{figure}

\end{enumerate}

We see that we have found two different ways to arrive at the same brane web. Note that the two webs in Figure \ref{web_2} are physically equivalent, while the ones in Figure \ref{web_1} are not (a Higgs branch flow has occurred).

\begin{figure}[h]
	\centering
	\includegraphics[height=2.5cm]{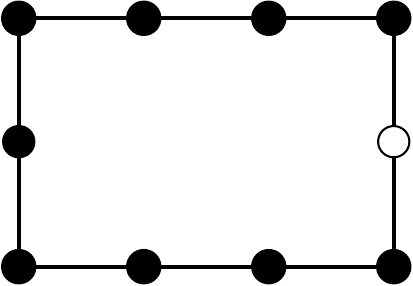}
\caption{GTP for the brane web in Figures \ref{web_1} and \ref{web_2}.}
	\label{GTP_1}
\end{figure}

It is possible to associate a cousin of the toric diagram for the case when multiple 5-branes end on the same 7-brane \cite{Benini:2009gi}. This object has been dubbed Generalized Toric Polygon (GTP) in the literature and is obtained by replacing a black dot with a white dot whenever it splits a segment corresponding to two 5-branes that are stuck together as in Figure \ref{GTP_1}, corresponding to the web in Figure \ref{web_2}. Note that in terms of the web, the effect of the s-rule due to the ending on 7-branes is to constrain the possible Coulomb branch deformations. Rules to account for that in terms of a ``poligonation'' of the GTP have been proposed in \cite{Bourget:2023wlb,Benini:2009gi}.

Returning to the relation to geometric engineering, the duality between M-theory on a CY$_3$ and a 5-brane web in Type IIB along the directions of the pinching of the torus only holds for webs without the 7-branes. In the case of webs with all external legs of multiplicity equal to 1, the external 7-branes can be neglected as they do not impose any (supersymmetry) constraint on the possible deformations. Thus, we can alternatively describe the corresponding SCFTs through a 5-brane web in Type IIB or as M-theory on the CY$_3$ whose toric skeleton coincides with the web. However, for generic multiplicities the question is: what does the GTP correspond to in the M-theory geometric engineering language? Progress in this direction was made in \cite{Bourget:2023wlb}: in the particular case where white dots appear only on one side of the GTP, it is possible to understand them by exploiting the reduction from M-theory to Type IIA. In the toric case, one side of the toric diagram corresponds to a (real) codimension-4 A-type du Val singularity, and after reduction on $S^1$ it appears as the D6 brane of Type IIA. In this setup, it is possible to give a nilpotent VEV to the stack of branes which are then forced to remain stuck together (a construction known as a T-brane \cite{Cecotti:2010bp}). After lifting back to M-theory, the conjecture is that the defining equations for the geometry look the same as for the toric diagram, except that fewer deformations are available.

In this work, we will argue that this is also the case for more generic GTPs, i.e. containing white dots in several sides as well as the interior. We will do so by exploiting mirror symmetry. We will see that while the geometric counterpart of the HW transition (analogous to Figure \ref{web_2}) is complicated to write down for the original M-theory CY$_3$ $\mathcal{M}$, it is straightforward to implement it for its mirror. This, in turn, allows us to identify the moduli that become frozen in general (analogous to Figure \ref{web_1}). 

\subsubsection{The $E_1$ example} 

\begin{figure}[h]
	\centering
	\includegraphics[height=3.8cm]{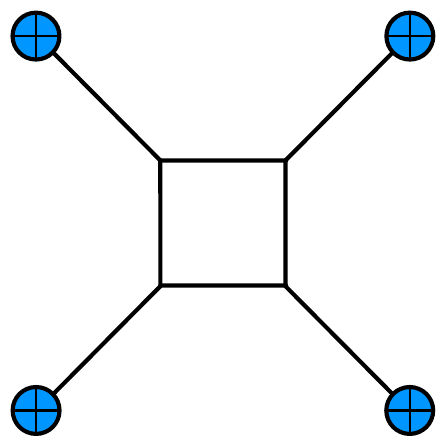}
\caption{Web for the $E_1$ theory.}
	\label{web_E1}
\end{figure}

Let us illustrate this discussion using the $E_1$ example. The corresponding web is shown in Figure \ref{web_E1}.

In this case, all external legs have multiplicity 1, which means that we are in the realm of standard toric diagrams. Indeed, the dual graph to Figure \ref{web_E1} is the resolved toric diagram of Figure \ref{toric_E1}. The proper Coulomb branch modulus corresponds to the vertical distance between the two parallel (1,0) 5-branes in the opened-up face; this is the mass of the fundamental string stretching between them. The extended Coulomb branch modulus corresponds to a deformation that moves the external legs horizontally, changing the mass of the D-string stretched between the two (0,1) 5-branes. Both proper and extended Coulomb moduli are K\"ahler moduli. They are distinguished because in one case we are not modifying the boundary conditions of the web (the face can open up while keeping the 7-branes fixed), while in the second one we are modifying them (moving some of the 7-branes transverse to the corresponding leg). Note also that while there are four external legs, two of their positions are fixed by charge conservation at every 5-brane junction, plus a third one from an overall translation, leaving us in total with two parameters for the extended Coulomb branch, which matches the geometric engineering result.

\begin{figure}[h]
	\centering
	\includegraphics[height=4cm]{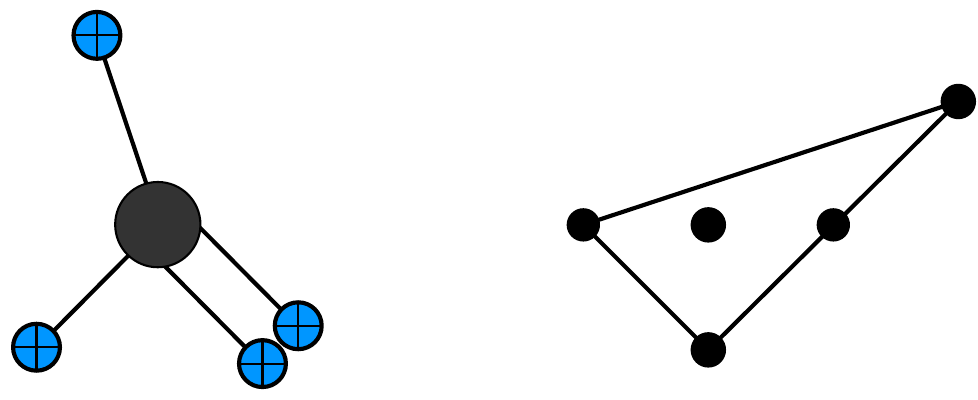}
\caption{Equivalent web for the $E_1$ theory after the Hanany-Witten transition.}
	\label{toric_web_mutation_E1}
\end{figure}

We may imagine now moving one of the 7-branes along the corresponding leg until it crosses the web. Without loss of generality (all 7-branes are equivalent), we choose the $[-1,1]$ one to cross. 
Moving the branch cut clock-wise, the $[1,1]$ 7-brane turns into a $[-1,3]$ 7-brane as it is swept by it. The resulting web is shown in Figure \ref{toric_web_mutation_E1}, together with the dual diagram, which is the toric diagram for $C(\mathbb{F}_2)$.

\subsection{Mirror construction}

Let us first focus on the realm of standard toric diagrams. Given a toric CY$_3$ $\mathcal{M}$ there is a well-known construction for its Hori-Vafa mirror partner $\mathcal{W}$ \cite{Hori:2000kt}. The toric geometry of the original $\mathcal{M}$ is encoded in its toric diagram, which can be regarded as a convex rational polytope $\Delta$ in the lattice $\mathbb{Z}^2$. Associated to $\Delta$ one can construct a Laurent polynomial by assigning to each point in $\Delta\cap \mathbb{Z}^2$ with coordinates $(m,n)$ the monomial $x^my^n$ (where $x,y\in\mathbb{C}^*$) with a generic coefficient $c_{(m,n)}$. The resulting polynomial is 
\begin{align}
    P(x,y)=\sum_{(m,n)\in \Delta\cap \mathbb{Z}^2} c_{(m,n)}x^my^n\,.
\end{align}
The mirror geometry is then constructed as a double fibration over a complex plane $\mathbb{C}$ parametrized by $w$ as
\begin{equation}\label{eq:mirror_general}
\mathcal{W}=\begin{cases} P(x,y)=w\,,\\ uv=w\,. \end{cases}
\end{equation}
There are three remarks. First, strictly speaking, this construction only works for $\Delta$'s with at least one internal point. These are called Fano polytopes, and in the remainder of this paper we will stick to that case, which implies that the 5d theory has rank greater or equal than 1. Second, note also that 
the Laurent polynomial $P(x,y)$ depends on the choice of origin of the lattice. While the choice of the origin is irrelevant for many applications and therefore hardly ever considered, it will become important in our discussion below, where it plays a crucial role in defining mutations of said polynomials. Finally, while it looks that we have as many free coefficients $c_{(m,n)}$ as points in the polytope, in fact we can always freely fix 3 of them, by using rescalings in $x$, in $y$, and an overall SL$(2,\mathbb{Z})$ transformation. We will often make use of this freedom.

Let us unpack the main features of the geometry of $\mathcal{W}$. For $w\ne 0$, $uv=w$ parametrizes a copy of $\mathbb{C}$, with an $S^1$ associated to the phase of the free complex coordinate (which can be taken to be $u$ or $v$); this $S^1$ collapses to zero size at $w=0$. In turn, for fixed $w$, the curve $P(x,y)-w=0$ defines a Riemann surface whose genus equals the number of internal points in $\Delta$. We can look for the critical points of $P$, i.e. for points $(x_I,y_I)$ such that

\begin{equation}
    \frac{\partial P}{\partial x}\Big|_{(x_I,y_I)}= \frac{\partial P}{\partial y}\Big|_{(x_I,y_I)}=0\,.
\end{equation}
They have the meaning that on top of every $w_I=P(x_I,y_I)$, there is an $S^1$ in the Riemann surface that pinches off. Thus, the segment $[0,w_I]$ on the $w$-plane connecting the 0 to $w_I$, together with the $S^1\times S^1$ associated to the two cycles above (one on the Riemann surface and pinching off at $w=w_I$ and the other one on the $uv$-plane and pinching off at $w=0$) defines a topological $S_I^3$. The number of critical points, and therefore of such $S^3_I$'s, is equal to the area of the toric diagram (in units of fundamental triangles). Moreover, this collection of spheres forms a basis of $H^3(\mathcal{W},\mathbb{Z})$, so the class of the $T^3$ toric fiber can be written as a formal linear combination of the $S_I^3$'s.

The fiber on top of $w=0$ is of special importance, which can be understood at an intuitive level by the fact that the $S^3_I$'s intersect on top of it. Let us denote by $\Sigma$ the corresponding Riemann surface $P(x,y)=0$. It contains the information of both the 5d theory as well as the corresponding BPS quiver, via the so called \emph{amoeba} and \emph{co-amoeba} projections \cite{Feng:2005gw}.

The amoeba projection of $\Sigma$ is defined as
\begin{align}
    \mathcal{A}_\Sigma = \left\lbrace \left(\log |x|,\log|y|\right), \text{ s.t. } P(x,y)=0\right\rbrace\,,
\end{align}
and it looks like a thickened version of the toric skeleton of the original CY$_3$ $\mathcal{M}$. Let us review how this comes about. Consider an edge of the toric diagram composed of a single segment connecting the points with coordinates $(a_1,b_1)$, $(a_2,b_2)$, so that the corresponding leg is, up to a sign, $(p,q)=(b_2-b_1,a_1-a_2)$. The equation $P=0$ can be written as $c_2 x^{a_2}y^{b_2}-c_1 x^{a_1}y^{b_1}=P'$, where $P'$ contains the contribution of all dots in the toric diagram other than the selected two. We can further write $c x^{-q}y^{p}-1=P''$, where $P''=P'/c_1 x^{a_1}y^{b_1}$ and $c=\frac{c_2}{c_1}$. Crucially, $P''$ can be regarded the contribution of all monomials as if the origin was set in the point $(a_1,b_1)$. Since the toric diagram is convex, all points lie to one side (depending on the particular choice of edge) of $(a_1,b_1)$. Thus, if we write $x=c^{\frac{1}{q}}\,t^p$, $y=t^q$, $P''$ is a polynomial with all positive/negative (depending on the side to which the rest of the points, as described before, lie) powers of $t$. Thus, in the appropriate limit $t\rightarrow 0,\infty$ $P''\rightarrow 0$, while the LHS obviously vanishes as well. Now, through the amoeba map above we have
\begin{align}
    \mathcal{A}_\Sigma \sim  \left(p\tau+p_0,q\tau\right), \qquad \tau=\log t\,,\qquad p_0=\frac{1}{p}\log c\,;
\end{align}
which corresponds to a infinite spike along the $(p,q)$ direction whose position is encoded in $c$. In particular, this shows that changes in $c$ map to changes of the positions of the external legs, that is, to mass deformations in the 5d SCFT. More generally, the coefficients $c_{(m,n)}$ in $P(x,y)$ are mapped to the extended Coulomb branch of the low energy theory. This is the familiar statement that mirror symmetry exchanges complex and K\"ahler moduli. 

Physically, the Riemann surface $\Sigma$ also makes an appearance as the analogue of the Seiberg-Witten curve for the 5d theory compactified on a circle. As described originally in \cite{Aharony:1997bh}, one writes $(x_5,x_6)$ for the coordinates in the plane of the web in Type IIB, and $(x_4,x_{10})$ for the coordinates on the $T^2$ where we compactify M-theory to make use of the duality between the two setups. Then the SW curve is given precisely by $P(x,y)=0$, where the coordinates are given by
\begin{align}\label{eq:coordinates_SW_curve}
    x = \exp\left[\frac{2\pi}{R}(x_5+ix_4)\right]\,,\qquad y=\exp\left[\frac{2\pi}{R}(x_6+ix_{10})\right]\,,
\end{align}
with $R$ the radius of the torus.\footnote{Since we have fixed $\tau=i$, which corresponds to the modular parameter of the M-theory torus, we are only left with specifying the base length $R$.} The various masses of BPS states of the 4d KK theory are computed using the usual SW technology, namely by integrating the differential $\lambda=\log x \, d(\log y)$ over the corresponding cycles.

The geometry of $\mathcal{W}$ also contains the information about the BPS quiver through the coamoeba projection of $\Sigma$. This is defined as the projection onto the angular parts of $(x,y)$, 
\begin{align}
    \mathcal{A}^\vee_\Sigma=\left\lbrace (\arg x,\arg y) \,,\text{ s.t. } P(x,y)=0\right\rbrace\,.
\end{align}
This projection lives on a $T^2$ and knows precisely how the $S^3_I$'s intersect one another on top of the fiber at $w=0$. Using the fact that the BPS quiver is identical to the fractional brane quiver for Type IIB D3-branes probing the original geometry, the BPS quiver appears through string dualities: the system of D3-branes probing $\mathcal{M}$ (in Type IIB) becomes, after three T-dualities, a system of D6-branes on $\mathcal{W}$ wrapping the $S^3_I$'s (in Type IIA). On $\Sigma$, the D6-branes look like 1-cycles, each of them surrounding a puncture associated to a leg of the amoeba, and intersecting one another according to the coamoeba. The combinatorics of this intersection pattern is captured by the brane tiling, which in turn specifies the $4d$ $\mathcal{N}=1$ theory living on the branes. We refer the reader to \cite{Feng:2005gw,Franco:2016qxh} for detailed discussions of this construction.

\subsubsection{The $E_1$ example} 

Let us revisit our trusty $E_1$ example. The toric diagram was depicted in Figure \ref{toric_E1}. From the associated Laurent polynomial, we find the mirror partner to $C(\mathbb{F}_0)$,
\begin{align}
\begin{cases}
        w = P(x,y) = c_0 + c_1 x + c_2 y + c_3 \frac{1}{x} + c_4 \frac{1}{y}\\
        w = uv
    \end{cases}\,,
\end{align}
where $u,v\in\mathbb{C}$ and $x,y\in \mathbb{C}^*$. As discussed after \eqref{eq:mirror_general}, we have the freedom to fix three of the complex coefficients $c_i=1$. In particular, we see that we have two complex structure moduli, as expected.

\begin{figure}[h]
	\centering
	\includegraphics[height=4.5cm]{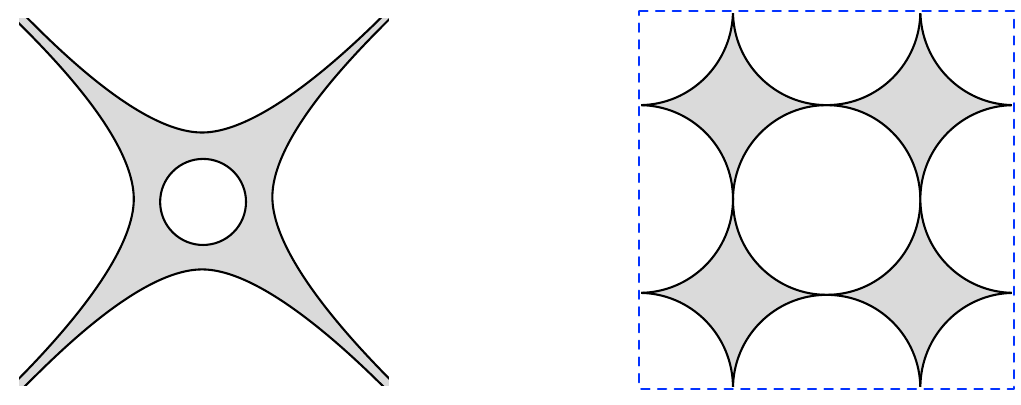}
\caption{Amoeba and coamoeba projections for the Riemann surface for $F_0$.}
	\label{amoeba_coamoeba_F0}
\end{figure}

The equation $P(x,y)=0$ is the Riemann surface above the origin of the $w$-plane. It is straightforward to determine its amoeba and coamoeba projections for specific values of $c_i$. Figure \ref{amoeba_coamoeba_F0} shows cartoons of the amoeba and coamoeba projections of $\Sigma$ for a choice of coefficients.\footnote{The coefficients in this example give rise to the so-called Phase 1 of $C(\mathbb{F}_0)$, given by Figure \ref{tiling_quiver_F0} and \eqref{W_F0_1}. Varying these coefficients, it is possible to obtain a qualitatively different coamoeba, which corresponds to Phase 2. In other words, the new coamoeba gives rise to the corresponding brane tiling or, equivalently, the quiver and superpotential. Detailed discussions of different choices of the coefficients, and the different toric phases (including specific analyses of $C(\mathbb{F}_0)$), can be found in \cite{Feng:2005gw,Franco:2016qxh}.}

We immediately see that the shape of the amoeba corresponds to the brane web in Type IIB. Even though the exact map between $c$'s and (extended) Coulomb branch parameters is relatively complicated, one can see that by varying the $c$'s it is possible to independently tune the size of the central hole (Coulomb branch VEV) and the horizontal distance between legs (mass parameter, in this case corresponding to inverse squared gauge coupling).

In the coamoeba the $S^3_I$'s are projected to $S^1$'s on $T^2$. These are wrapped by D6-branes, and we see that their pattern of intersections reproduces the brane tiling shown in Figure \ref{tiling_quiver_F0}.

\section{Hanany-Witten transitions and polytope mutations}\label{sec:HWandMutation}

This section contains the main point of this paper, the implementation of the Hanany-Witten transition of the brane web in terms of the geometry, which is particularly easily visualized in the mirror $\mathcal{W}$. The corresponding operation is known in the mathematical literature as a \emph{polytope mutation}. The connection between Hanany-Witten transitions an polytope mutations was already noted and investigated in \cite{Franco:2023flw}. In this paper, however, we consider a refined version of polytope mutation which, among other things, also specifies the transformation of the Laurent polynomial. This prescription enables the mapping of moduli across the transition, allowing the determination of instances when they become frozen.

\subsection{Hanany-Witten in the mirror}

\begin{figure}[h]
	\centering
	\includegraphics[height=6cm]{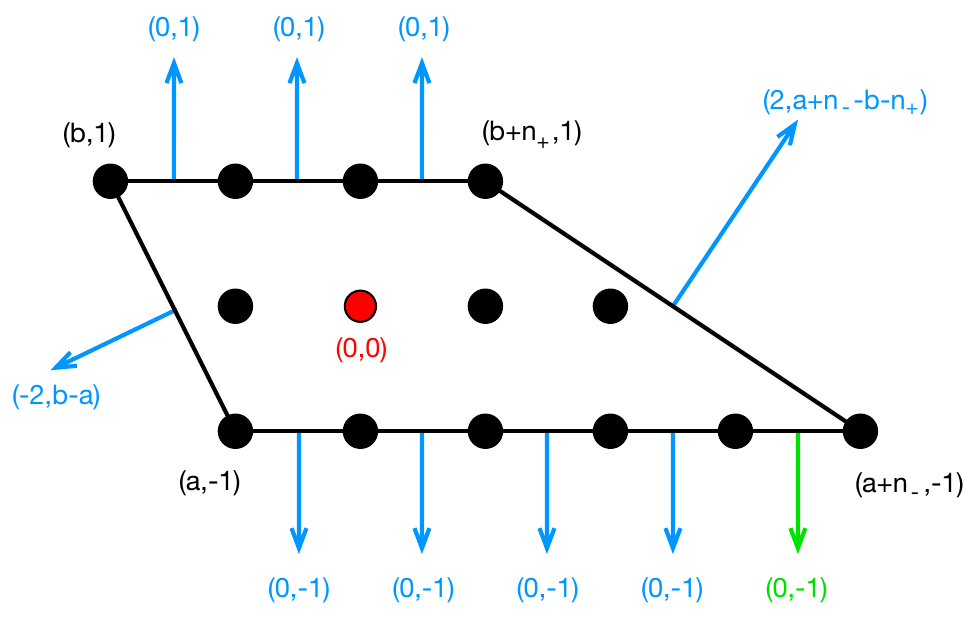}
\caption{Polytope associated to the polynomial \eqref{eq:Laurent_example_length_2} for $a=-1$, $b=-2$, $n_+=3$ and $n_-$=5. The origin is shown in red. We show the vectors normal to every edge on the boundary.}
	\label{polytope_mutation_HW_1}
\end{figure}

Consider a polytope $\Delta$ and its associated Laurent polynomial $P(x,y)$. As a warm-up, let us illustrate our construction with a four-sided polytope with two parallel sides such that the distance between them is 2 (in lattice units). Without loss of generality, we can use an appropriate SL$(2,\mathbb{Z})$ transformation such that one of the parallel edges is on the $x$ axis and the other one is at $y=2$, as illustrated in Figure \ref{polytope_mutation_HW_1}.

Picking the origin of the lattice in the interior of $\Delta$, i.e. with $y=1$, the corresponding polynomial takes the following form
\begin{align}\label{eq:Laurent_example_length_2}
    P(x,y) =& \frac{1}{y}P_{-1}(x)+y^0 P_0(x) + y P_1(x) \\
    =&\frac{x^a}{y}\prod_{i=1}^{n_{-}} (x-x_i) + P_0(x) + y \, x^b\prod_{j=1}^{n_+} (x-\widetilde{x}_j)\,.
\end{align}
The exponents $a$ and $b$ resulting from the factorization of $P_{\pm 1}$ can be either positive or negative, and they depend on the precise choice of origin along the $x$ axis. Figure \ref{polytope_mutation_HW_1} shows the general form of $\Delta$.

As discussed in the previous section, this polynomial specifies a geometry $\mathcal{W}$ whose amoeba projection (of the Riemann surface $\Sigma$) has vertical asymptotes along the bottom edge at positions $x_i$ and along the top edge at positions $\widetilde{x}_j$. The dual IIB brane web will have $n_{-}$ semi-infinite  NS5-branes at the bottom and $n_+$ NS5-branes at the top, respectively at positions 
\begin{align}
    x_{5,i} = \frac{R}{2\pi}\log x_i\,,\qquad \widetilde{x}_{5,j} = \frac{R}{2\pi}\log \widetilde{x}_j\,,
\end{align}
after using \eqref{eq:coordinates_SW_curve}.

Let us now consider the following change of coordinates:
\begin{align}\label{eq:mutation_1}
    (x,y)\to (x,(x-x_1)y)\,.
\end{align}

\begin{figure}[h]
	\centering
	\includegraphics[height=6cm]{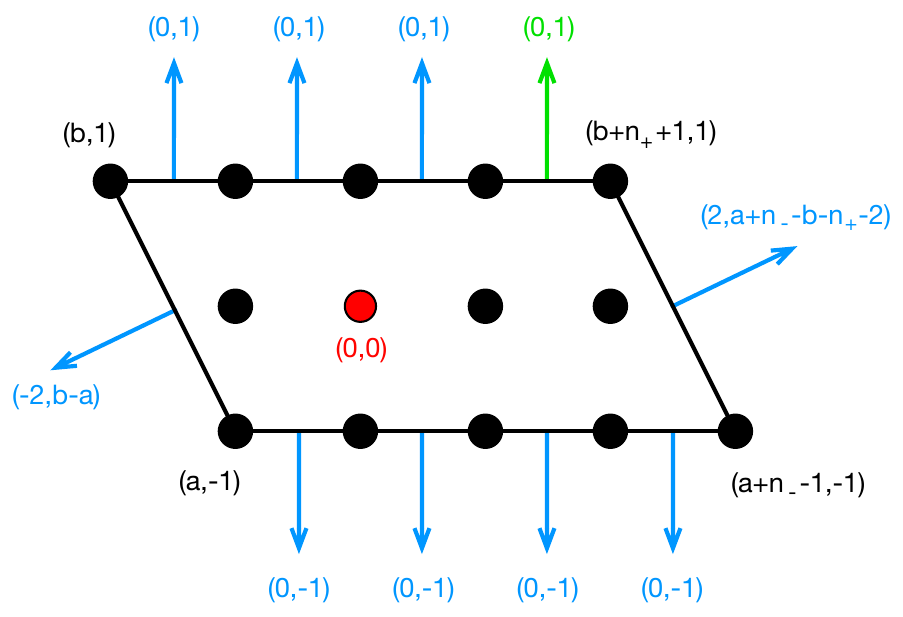}
\caption{Mutated polytope obtained from Figure \ref{polytope_mutation_HW_2} by the change of variables in \eqref{eq:mutation_1}. The mutation implements a Hanany-Witten transition in the dual web.}
	\label{polytope_mutation_HW_2}
\end{figure}

It is immediate to see what are the consequences for the Laurent polynomial \eqref{eq:Laurent_example_length_2}. The $(x-x_1)$ factor in $P_{-1}(x)$ cancels, and it appears in $P_1(x)$ instead. We obtain
\begin{align}
    P(x,y)\to \frac{x^a}{y}\prod_{i=2}^{n_{-1}}(x-x_i) + P_0(x) + y\, x^b\, (x-x_1)\prod_{j=1}^{n_1}(x-\widetilde{x}_j)\,.
\label{polytope_HW_mutation_1}
\end{align}
The resulting polytope is shown in Figure \ref{polytope_mutation_HW_2}. In terms of the dual Type IIB brane web, we have sent the external leg at position $x_1$ from the bottom to the top of the brane web. Moreover, this automatically changes the slope of the edge at the rightmost side of the polytope.\footnote{While the explicit example in the figure was chosen such that before and after mutation both of the sides of $\Delta$ that are not parallel to the $x$-axis consist of a single edge (namely they do not cross over intermediate lattice points), our discussion extends without changes to the case in which these sides contain multiple edges. Moreover, the fact that the two lateral edges end up being parallel after the mutation is just a non-generic feature of this specific example.} This is precisely the effect of sweeping the monodromy in the Hanany-Witten transition resulting from moving the 7-brane at the end of the leg across the web.

We can use the same type of coordinate transformations to implement more general HW transitions. For example, in order to send one of the legs from the top to the bottom of the web, we would use the transformation
\begin{align}\label{eq:mutation_2}
    (x,y)\to (x,\frac{y}{x-\widetilde{x}_j})\,.
\end{align}
Similarly, if we want to send several 7-branes across the web, we can employ the transformation
\begin{align}\label{eq:mutation_3}
    (x,y)\to(x,p(x)\, y)\,,
\end{align}
with the restriction that $p(x)$ is a divisor of $P_{-1}(x)$. This requirement ensures that the corresponding monomials cancel, and physically has the meaning that we are picking the subset of the legs that appear as factors in $P_{-1}(x)$.

An interesting new phenomenon arises when the length of the polytope $\Delta$ in the direction of $y$ is greater than 2. Consider, for example, an example of height 3, and pick the origin such that the Laurent polynomial is
\begin{align}
    P(x,y) = \frac{1}{y} P_{-1}(x) + P_0(x) + y P_{1}(x) + y^2 P_2(x)\,,
\end{align}
and take $(x-x_1)$ a prime factor of $P_{-1}(x)$. 

Then the coordinate transformation $(x,y)\to(x,(x-x_1)y)$ sends $P_2(x)\to P_2(x)(x-x_1)^2$. We observe that we have fewer free complex coefficients than the naive counting from the number of monomials. In the dual Type IIB picture, this means that two of the external legs created by the HW transition are fixed to be together at the same position $x_1$. We interpret this as the fact that two 5-branes end on the same 7-brane, and correspondingly, that we have a white dot in the corresponding GTP. 

Having a height 3 polytope, allows us to pick different vertical positions for the origin. More generally, the distance between the origin and the edge we want to mutate can take different values. We now discuss the important role of the choice of origin. Returning to the example at hand, depending on the choice of origin, we can shift the entire polytope vertically, i.e. shift the $y$ powers in the Laurent polynomial. In particular, we can make an alternative choice for which $P(x,y)$ becomes
\begin{align}\label{eq:example_mutation_white_dot_2}
     P(x,y) = \frac{1}{y^2}P_{-2}(x)+\frac{1}{y} P_{-1}(x) + P_0(x) + y P_{1}(x)\,.   
\end{align}
Comparing to and \eqref{polytope_HW_mutation_1} and Figure \ref{polytope_mutation_HW_1}, we have moved the polytope downwards by one lattice unit in the $y$ direction. Equivalently, we have picked the origin to be one unit further from the lower edge. Now, if we want to send $y\to (x-x_1)y$ we also need to require that $(x-x_1)^2$ divides $P_{-2}(x)$ to ensure that the desired cancellations happen and the final result is also a Laurent polynomial. This means that before the coordinate transformation, we must have two legs of the web frozen together, which become just one leg after the transition. This is the first instance where we notice that the choice of origin of the lattice is related to whether the white dot is present in the GTP before or after the HW transition, a point to which we will return later.

The transformations \eqref{eq:mutation_1}, \eqref{eq:mutation_2}, \eqref{eq:mutation_3} are examples of so called \emph{mutations} of Laurent polynomials, which in turn implies mutations of their Newton polytopes. The key point is that the polytope mutation is implemented by a change of variables in the corresponding Laurent polynomial. As such, it provides a more refined description than the mere mutation of the points in the polytope. These mutations have been discussed in the pure mathematics literature quite extensively starting with \cite{GalkinUsnich}. As we will illustrate in examples below, mutations on edges that are not horizontal can be obtained directly by appropriate changes of variables or, equivalently, by first applying an SL$(2,\mathbb{Z})$ transformation to turn the edge under consideration horizontal. In the next subsection, we will review some of the mathematical terminology and results, so that we can then import them to Physics.

\subsubsection{The $E_1$ example} 

\begin{figure}[h]
	\centering
	\includegraphics[height=3cm]{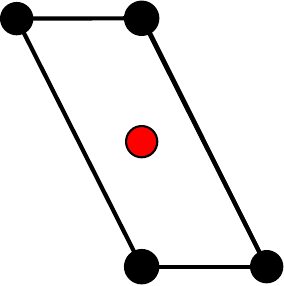}
\caption{Polytope for the $E_1$ theory obtained from Figure \ref{toric_E1} via an SL$(2,\mathbb{Z})$ transformation. The origin is indicated in red.}
	\label{toric_E1_2}
\end{figure}

Let us illustrate the previous discussion in the example of the $E_1$ theory. Instead of starting with the polytope in Figure \ref{toric_E1}, it is slightly more convenient to perform a global SL$(2,\mathbb{Z})$ transformation to align one side of the polytope with the $x$ axis, as shown in Figure \ref{toric_E1_2}.

\begin{figure}[h]
	\centering
	\includegraphics[height=3cm]{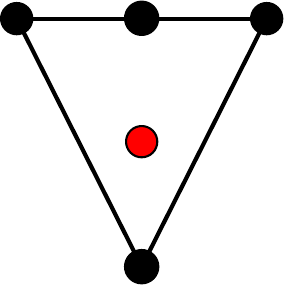}
\caption{Polytope obtaines from Figure \ref{toric_E1_2} by mutating it on the lower horizontal edge.}
	\label{toric_F2_2}
\end{figure}

The corresponding Laurent polynomial reads
\begin{equation}
    P(x,y)=\frac{1}{y}(x-c_1)+1+y\,\left(\frac{1}{x}-\frac{1}{c_2}\right)\,,
\end{equation}
where we have already fixed three of the coefficients, and the remaining two are directly identified with the position of the external legs of the web. Now we perform the following mutation,
\begin{equation}\label{eq:mutation_E1_example}
    (x,y)\rightarrow(x,(x-c_1) y)\,,
\end{equation}
and find
\begin{equation}\label{eq:mutation_E1_bis}
    P(x,y)\rightarrow \frac{1}{y}+1+y\,\left(\frac{1}{x}-\frac{1}{c_2}\right)(x-c_1)\,,
\end{equation}
whose polytope is shown in Figure \ref{toric_F2_2}, which corresponds to $C(\mathbb{F}_2)$ (up to an overall rotation, it is the one in Figure \ref{toric_web_mutation_E1}). In particular, note once again that the mutation of the Laurent polynomial automatically knows about the action due to sweeping the monodromy cut of the 7-brane across half of the web --this is a completely generic fact. In this example, the HW transition does not generate a white dot in the toric diagram; accordingly, none of the moduli in \eqref{eq:mutation_E1_bis} are frozen.
 
\subsection{Polytope mutations}

As mentioned above, HW transitions translate, in the mirror, to the notion of mutation defined in the mathematical literature \cite{GalkinUsnich}. The starting point is a toric diagram in the standard sense --that is, with no decoration of white dots-- which can be regarded as a rational polytope $\Delta$ in $\mathbb{Z}^2$. We assume that the toric diagram contains the origin in its interior, which in turn implies that it has at least one internal point. This amounts to restricting to 5d SCFTs of rank greater or equal than 1. In the mathematical literature, it is also often required that the vertices are primitive, which means that their coordinate vectors consist of coprime numbers.\footnote{Note that vertices being primitive or not depends on the choice of origin. We will see examples of this below.}. However, we will not make this requirement here. Given one such $\Delta$, we can always write the corresponding Laurent polynomial $P(x,y)$ as a Laurent polynomial in $y$, whose coefficients are Laurent polynomials in $x$. Since the origin is in the interior of $\Delta$, $P(x,y)$ contains a finite number of terms with negative and positive powers for both $y$ and $x$. We can write
\begin{equation}
\label{eq:Porganized}
    P(x,y)=\sum_{n=-N_-}^{N_+} P_n(x) y^n\,,
\end{equation}
where $N_-$ and $N_+$ are the maximum negative and positive powers of $y$, respectively. $P_n(x)$ are Laurent polynomials in $x$, whose degrees need to satisfy that the corresponding Newton polygon is convex. A mutation is then a birrational transformation of the form \cite{GalkinUsnich}
\begin{equation}
\label{eq:mutationmath}
    \mu:\,(x,y)\rightarrow (x,\alpha(x)\,y)\,,
\end{equation}
where $\alpha(x)$ is a Laurent polynomial such that $\alpha^i(x)$ divides $P_{-i}(x)$. Clearly, the transformation in \eqref{eq:mutation_1} or its generalization \eqref{eq:mutation_3} are particular cases of \eqref{eq:mutationmath}. Thus, the physical avatar of the mathematical notion of mutation is crossing 7-branes in a 5-brane web. 

\begin{figure}[h]
	\centering
	\includegraphics[height=6cm]{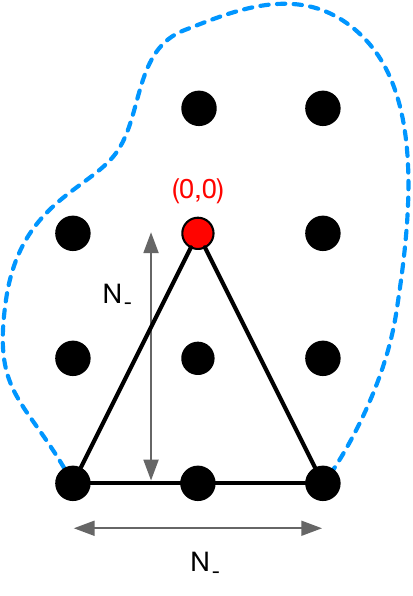}
\caption{The triangle selected in the mutation. The dashed line represents the rest of the diagram, whose precise shape is not relevant, as long as it is convex. The red dot represents the origin.}
	\label{primitive_cone}
\end{figure}

Note that, in order for $\alpha^i$ to divide $P_{-i}$, some conditions must be met. First, there are requirements on the degree of the polynomials $P_i$. Let us consider the case of $\alpha(x)$ a polynomial of degree 1. In that case, it is clear that $P_{-i}$ must be at least a polynomial of degree $i$ in $x$. Moreover, the coefficients of the polynomial must be tuned so as to have the relevant number of common roots, as was discussed already around \eqref{eq:example_mutation_white_dot_2}. All in all, this selects a triangle of points ($n$ points at lattice distance $n$ from the origin). In total, the triangle will have base $N_-$ and height $N_-$ with respect to the origin, as depicted in Figure \ref{primitive_cone}.

Then, the action of mutation is to insert a triangle on the the other side with respect to the origin, only that this time with its base of length and height are equal to $N_+$. Finally, if $\alpha$ is a polynomial of arbitrary degree $d$, then we can always factor it, and a similar argument would hold for each of the $d$ monomials. That is, we would end with $d$ triangles of equal base and height $N_-$ before the mutation, that would transform into $d$ triangles of equal base and height $N_+$ after it. This allows us to translate the action of the mutation directly in terms of the polytope, without the need to go through the Laurent polynomial as an intermediate step.

The way this is made precise in the mathematical literature is by introducing the notion of a \emph{primitive T-cone}. A primitive T-cone is a cone formed by the origin and a an external side (or a collection of edges on a side) of lattice length $N$ which is at lattice distance $N$ from the origin. The triangles in Figure \ref{primitive_cone} and discussed above are examples of this construction. A primitive T-cone of base $N$ is naturally identified with a 7-brane with $N$ 5-branes ending on it. Let us stress that the definition of primitive T-cone gives a physical meaning to the position of the origin of the polytope, as the height, and therefore the base length of the corresponding primitive T-cone, depends on it. This phenomenon was already discussed above in terms of the mutation of Laurent polynomials. In pure polytope language, the mutation amounts to removing one primitive T-cone from one side of the polytope and adding a new primitive T-cone to the other side. Since generically the other side of the polytope is at a different lattice height $N_+\ne N_-$, the resulting primitive T-cone will generically be of different length $N_+$, which is the manifestation of the Hanany-Witten effect. In fact, one can be fully precise and describe the effect of the mutation combinatorially (see \textit{e.g.} \cite{Higashitani:2019vzu}), precisely finding \eqref{HWtransf}. This shows, as anticipated in \cite{Franco:2023flw}, that the mathematical notion of mutation precisely coincides with the HW transitions in brane webs.

Note that, with a choice of origin such that $N_+$ is much bigger than $N_-$ (or vice versa), the initial polytope $\Delta$ and its mutation, which we will denote $\mu(\Delta)$, may differ a lot in size (in plain words, one of them may have much more points than the other, precisely as a consequence of the HW effect). This is not a problem precisely due to the fixing of the various complex coefficients $c_{(n,m)}$ when connecting both sides of the mutation. As observed above, in the geometry of the mirror $\mathcal{W}$, this is understood as the freezing of some of the complex moduli, while in the original CY$_3$ $\mathcal{M}$ it translates as a freezing of the K\"ahler moduli; precisely in such a way as to preserve the extended Coulomb branch of the 5d SCFT.

Importantly, this freezing of moduli does not affect the functional dependence of $P(x,y)$ on the variables $x$ and $y$, and thus the geometry of the mirror $\mathcal{W}$. Correspondingly, we argue that the geometry $\mathcal{M}$ corresponding to a GTP is the same as for the standard toric diagram (namely, with no white dot decoration) except that some of its K\"ahler moduli are frozen. This is consistent with the proposal in \cite{Bourget:2023wlb}, and we will come back to it in Section \ref{sec:periods}.

\subsection{Mutation invariants}\label{sec:invariants}

In the mathematical literature, an important question is how to determine when two polytopes can be connected via a sequence of mutations. To that end, three invariant quantities have been defined: the \emph{singularity content}, the \emph{classical period} and the \emph{Hilbert series} \cite{akhtar2012minkowski}. We will not discuss the first invariant, except to mention that it corresponds to how each of the edges looks as a codimension 4 singularity (see \cite{2014arXiv1401.5458A} for more details). The second two, to which we now turn, can be computed from the Laurent polynomial $P(x,y)$ associated to the polytope; we should note that the choice of origin will be relevant in general. Lastly, one word on notation: we will often denote either the period or the Hilbert series by the name of the CY$_3$ variety whose toric diagram is $\Delta$. In what follows, we will always assume that the definition of the polytope $\Delta$ already incorporates the choice of origin.

\begin{itemize}
\item The (classical) period is defined in terms of the Laurent polynomial $P(x,y)$ associated to $\Delta$ as
    \begin{equation}
        \pi_{\Delta}(t)=\frac{1}{(2\pi i)^2}\oint_{|x|,|y|= 1}\frac{dx\,dy}{x\,y}\,\frac{1}{1-t\,P(x,y)}\,,
    \end{equation}

where the variables $x,y\in\mathbb{C}^*$. In the context at hand, it has been argued that the classical period defined above coincides with the quantum period and it is a generating function for Gromov-Witten invariants of the original Calabi-Yau $\mathcal{M}$ \cite{2012arXiv1212.1722C,2022arXiv221007328C}. Again we remark that $\pi_{\Delta}(t)$ generically depends on the choice of origin of the polytope.

\item The \textit{Hilbert series} of the variety $\mathcal{M}$. Since we can label the variety by the polytope $\Delta$, we will write ${\rm Hilb}_{\Delta}(t)$. It turns out that this can be easily computed as the Ehrhart series of the dual polytope $\Delta^{\circ}$ \cite{akhtar2012minkowski}. The dual polytope is defined as
\begin{equation}
    \Delta^{\circ}=\{u\in \mathbb{Q}^2\,/\,u\cdot v\geq -1\,,\forall\, v\in\Delta\}\,.
\end{equation}
In practice, it is sufficient to consider the inequalities obtained when $v$ are the vertices of $\Delta$. Note that the dual polytope need not be integral. Then,
\begin{equation}\label{eq:def_ehrhart}
{\rm Hilb}_{\Delta}(t)={\rm Ehr}_{\Delta^{\circ}}(t)=\sum_{n=0}|n\Delta^{\circ}\cap \mathbb{Z}^2|\,t^n\,.
\end{equation}
In words, $|n\Delta^{\circ}\cap \mathbb{Z}^2|$ is the number of integral lattice points in the $n$ times enlarged copy of $\Delta^{\circ}$. Once again, note that this quantity depends on the choice of origin inside the polytope. 

\end{itemize}

These quantities are invariant under mutation $\Delta\rightarrow \mu(\Delta)$, namely
\begin{equation}
    \pi_{\Delta}(t)=\pi_{\mu(\Delta)}(t)\,,\qquad {\rm Hilb}_{\Delta}(t)={\rm Hilb}_{\mu(\Delta)}(t)\,.
\end{equation}

An important remark is that the period $\pi_\Delta(t)$ depends in general on the coefficients $c_i$, and therefore the fact that it is invariant under mutation is yet another indication that for generic toric geometries the complex moduli becoming frozen is the appropriate interpretation of the white dot in the GTP.

Moreover, it has been conjectured \cite{2012SIGMA...8..047I,2015arXiv150105334A,2021RSPSA.47710584C} (see also \cite{2022arXiv221007328C}) that whenever the period of two polytopes $\Delta_1$ and $\Delta_2$ coincide, then: 1) there exists a sequence of mutations connecting the two polytopes and 2) there exists a flat fibration of geometries interpolating between $\mathcal{M}_{\Delta_1}$ and $\mathcal{M}_{\Delta_2}$. In physical terms, the first result can be used to determine when two brane webs, or their associated geometries, can be connected via a collection of HW transitions or their corresponding mutations. On the other hand, the second result implies that there exists a geometry (in general non-toric) that continuously describes the HW transition.

\subsubsection{The $E_1$ example}

In the $E_1$ example, the period associated to the polytope corresponding to $C(\mathbb{F}_0)$ is
\begin{align}\label{eq:period_examples_E1}
    \pi_{C(\mathbb{F}_0)}(t)=\int \frac{dx\, dy}{x\, y}\frac{1}{1-t\left[\frac{1}{y}(x-c_1) + 1 + y\left(\frac{1}{x}-\frac{1}{c_2}\right)\right]}\,.
\end{align}
The integral can be computed as a series expansion in $t$ by expanding the geometric series in the integrand. The first few orders in the series are
\begin{align}
    \pi_{C(\mathbb{F}_0)}(t)=1+t+\left(3+2\frac{c_1}{c_2}\right)t^2+O(t^3)\,.
\end{align}
The fact that the mutation \eqref{eq:mutation_E1_example} preserves the period follows trivially from implementing it as a change of variables in the integral \eqref{eq:period_examples_E1}: the change in the Haar measure is absorbed by the Jacobian of the change of variables. In fact, the same argument also applies for generic mutations of Laurent polynomials; instead, it is the reverse implication that is highly non-trivial.

The second invariant can be computed as follows. The polytopes $\Delta_{C(\mathbb{F}_0)}$ and $\Delta_{C(\mathbb{F}_2)}$ have vertices $\{(1,-1),(0,-1),(-1,1),(0,1)\}$ and $\{(0,-1),(-1,1),(1,1)\}$, respectively.\footnote{Here we consider the polytopes for these geometries given in Figures \ref{toric_E1_2} and \ref{toric_F2_2}, which are related by mutation.} Therefore, the dual polytopes are
\begin{align}
  \Delta_{C(\mathbb{F}_0)}^\circ &= \{(a,b)\in\mathbb{Q}^2\text{ s.t. } -1\le b\le 1\,,\, b-1\le a\le b+1\}\,,\\  
    \Delta_{C(\mathbb{F}_2)}^\circ &= \{(a,b)\in\mathbb{Q}^2\text{ s.t. } b\le 1\,,\, -b-1\le a\le b+1\}\,.
\end{align}

From this, it is straightforward to scale the size of the dual polytopes and count the number of internal integral lattice points at each $n$ as in \eqref{eq:def_ehrhart}, finding that they are indeed the same and equal to
\begin{equation}
\label{eq:HilbE1}
    {\rm Hilb}_{C(\mathbb{F}_0)}={\rm Hilb}_{C(\mathbb{F}_2)}=\frac{1+6t+t^2}{(1-t)^3}\,.
\end{equation}

\section{5d SCFTs, GTPs, and the period}\label{sec:periods}

We have seen that there is a deep connection between the mathematical theory of mutations and the physics of 5d SCFTs and their engineering in String Theory. Let us make a quick summary of the discussion above:

\begin{itemize}
    \item There are two ways in which to reach a brane web corresponding to a GTP: a Hanany-Witten transition (as in Figure \ref{web_1}) and a Higgs branch flow (as in Figure \ref{web_2}). 
    
    \item The HW transition corresponds to a mutation of the geometry. One can easily keep track of the various moduli by looking at the geometry of the mirror $\mathcal{W}$ and, in particular, see that the extended Coulomb branch of the 5d SCFT is preserved.

    \item The Higgs branch flow corresponds to freezing some of the moduli in M-theory geometric engineering (complex moduli in the mirror $\mathcal{W}$, K\"ahler moduli in the original $\mathcal{M}$) while keeping the geometry otherwise identical. This process changes the extended Coulomb branch of the 5d theory, as expected. 
\end{itemize}

Now, we turn to explore some consequences of the mathematical work on mutations in our physical setup. Here, we will discuss three items:

\begin{enumerate}
    \item The period $\pi_\Delta(t)$ can be used to efficiently classify 5d SCFTs coming from brane webs. In particular, it allows us to determine when two \emph{a priori} different looking webs can be related by a sequence of HW transitions.

    \item When two toric geometries are related by a mutation, it is possible to build a flat family of geometries interpolating between the two. This corresponds to the geometric version of the HW transition in the original M-theory Calabi-Yau $\mathcal{M}$.

    \item The mirror geometry $\mathcal{W}$ automatically encodes the Seiberg-Witten curve of the 5d theory on a circle. This can be used to extract the effect of the frozen moduli on the low energy physical observables \cite{Kim:2014nqa}.
\end{enumerate}

We should remark that in this section we will be considering simple examples as proof of concept of our methods. It would be very interesting to investigate these problems in more depth, which we postpone for future work.

\subsection{The period and the classification of brane webs}

Due to the constraints imposed by supersymmetry in 5 dimensions, the exploration of the full landscape of theories appears to be an achievable goal, at least for low rank \cite{Jefferson:2018irk,Kim:2020hhh}. While most efforts have tackled this problem from a geometric engineering point of view, it is also possible to attempt a similar program in terms of brane webs \cite{Arias-Tamargo:2022qgb}. In this context, the primary challenge lies in discerning whether two brane webs correspond to distinct 5d SCFTs or if there exists a sequence of duality transformations and HW moves that transforms one into the other. Several invariants under these transformations have been defined in order to tackle this problem \cite{DeWolfe:1998eu}. They are the \emph{total monodromy} of the brane web and the \emph{asymptotic charge invariant}. The total monodromy is the product of the monodromies associated to each 7-brane of the web. In the notation of Section \ref{sec:Brane_webs_GTPs},
\begin{align}
    M_{\rm tot} = \prod_{\ell} M_\ell\,.
\end{align}
The asymptotic charge invariant $\mathcal{Q}$ is
\begin{align}
    \mathcal{Q} = \text{gcd}\{ \langle \ell_i\, | \,\ell_j \rangle \,,\,\forall\, i,j \}\,.
\end{align}

\begin{figure}[h]
	\centering
	\includegraphics[height=4cm]{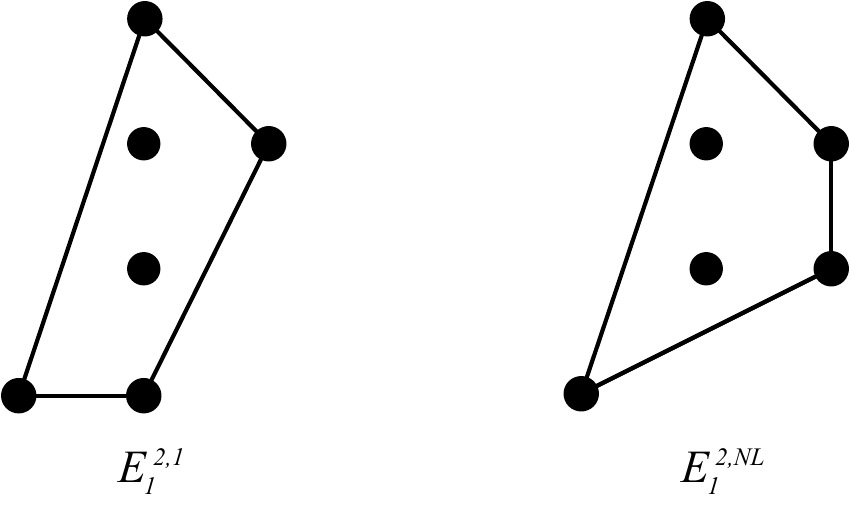}
\caption{The two webs to compare, taken from \cite{Saxena:2020ltf}.}
	\label{toric_to_compare}
\end{figure}

Interestingly, the fact that these quantities are identical for two webs is necessary but not sufficient for them to define the same low-energy 5d theory. This raises a puzzle already at rank 2 \cite{Arias-Tamargo:2022qgb}, namely that the total monodromy and asymptotic charge invariant do not completely specify the SCFT. The specific exampled discussed in that reference were the webs labelled by (c) and (e) in Figure 3 of \cite{Saxena:2020ltf} and shown in Figure \ref{toric_to_compare}. These two webs share the same classifiers, yet they are expected to give rise to different low energy theories.

One immediately sees that the period $\pi_\Delta(t)$ seems to be precisely the right tool to perform this task. As we have discussed above, there is strong evidence that whenever the periods coincide there exists a sequence of mutations connecting the corresponding Laurent polynomials, and it has been conjectured that this should indeed be true in general  \cite{2012SIGMA...8..047I,2015arXiv150105334A,2021RSPSA.47710584C}. That is, it is to be expected that the identity of periods is a necessary and sufficient condition for the equivalence of the corresponding polytopes under mutation (and hence of the associated varieties). In physics terms, this would allow to discriminate whether or not two webs describe the same 5d theory.

Indeed, one can check that the two polytopes in Figure \ref{toric_to_compare} give rise to different periods. The computation of the period as a series expansion is straightforward, however two issues need to be taken into account: the choice of origin (since this example has two internal points) and a possible non-trivial map of the moduli (the coefficients $c_{(n,m)}$) from one polytope to the other. Here, we proceed by brute force. Namely, we consider both possible choices of origin and check that there is no possible matching of the moduli such that the periods are equal.

Let us denote $P_c^i(x,y)$ and $P_e^i(x,y)$ the corresponding Laurent polynomials for the two polytopes, where $i=1,2$ labels the possible choices of origin:
\begin{align}
    P_c^1(x,y) &= C_0 + C_1 y + C_2 y^2 + C_3 xy + C_4\frac{1}{y} + C_5 \frac{1}{xy}\,,\\
    P_c^2(x,y) &= D_0 + D_1 y + D_2x + D_3 \frac{1}{y} + D_4 \frac{1}{y^2} + D_5 \frac{1}{x y^2}\,,\\
    P_e^{1}(x,y) &= E_0 + E_1 y+ E_2 y^2 + E_3 x +E_4 x y+E_5\frac{1}{xy}\,,\\
    P_e^{2}(x,y) &= F_0 + F_1 y+ F_2 x + F_3\frac{1}{y}+F_4\frac{x}{y}+F_5\frac{1}{xy^2}\,.
\end{align}
Note that we have chosen not to fix any values of the complex coefficients. The first few orders of the periods for each of these polynomials read
\begin{align}
    \pi_{c,1}(t) = & 1 + C_0 t +  \left(C_0^2+2 C_1 C_4+2 C_3
   C_5\right) t^2 +\left(C_0^3+6 C_0 (C_1 C_4+C_3 C_5)+3 C_1
   C_4^2\right) t^3 \nonumber\\
  & + \left(C_0^4+12 C_0^2 (C_1 C_4+C_3
   C_5)+12 C_0 C_1 C_4^2+6 C_1^2 C_4^2+24 C_1
   C_3 C_4 C_5+6 C_3^2 C_5^2\right)  t^4 + O(t^5)\,, \\
    \pi_{c,2}(t)= & 1 + D_0 t +  \left(D_0^2+2 D_1 D_2\right) t^2 
    + \left(D_0^3+6 D_0 D_1 D_2+3 D_1^2 D_2\right) t^3 \nonumber\\
 &+  \left(D_0^4+12 D_0^2
   D_1 D_2+12 D_0 D_1^2 D_2+12 D_1^2 D_2
   D_2+6 D_1^2 D_2^2\right)t^4 + O(t^5)\,,\\
    \pi_{e,1}(t) = & 1 + E_0 t+ \left(E_0^2+2 E_4 E_5\right) t^2 + \left(E_0^3+6 E_0 E_4 E_5+6 E_1 E_3
   E_5\right) t^3  \nonumber\\
   & + \left(E_0^4+12 E_0^2 E_4 E_5+24 E_0 E_1 E_3
   E_5+6 E_4^2 E_5^2\right) t^4 + O(t^5) \,,\\
    \pi_{e,2}(t) = & 1 + F_0 t + \left(F_0^2+2
   F_1 F_3\right) t^2 + \left(F_0^3+6 F_0 F_1 F_3\right) t^3 \nonumber\\
   &+ \left(F_0^4+12 F_0^2 F_1 F_3+6
   F_1^2 \left(2 F_2 F_5+F_3^2\right)\right) t^4 + O(t^5)\,.
\end{align}

Already at fourth order one can explicitly check that there is no map between the coefficients of $P_c(x,y)$ and $P_e(x,y)$ (allowing for both possible choices of origin) such that their periods are equal. This shows that indeed these two polytopes are not related by a mutation, or in other words the two brane webs are not related by a Hanany-Witten move. Hence, they define two different 5d SCFTs.

This example is a proof of concept of how the periods associated to the polytopes might be used for classifying 5d SCFTs. Even though we will not pursue this further here, one could consider proceeding in a systematic fashion by e.g. listing the possible brane webs of a given rank, or a given number of external legs (in the same spirit of \cite{Arias-Tamargo:2022qgb}) and computing $M_{tot}$, $\mathcal{Q}$ and $\pi(t)$ to discern if they define the same low energy theory.

\subsection{Hanany-Witten and geometric deformations}

Having identified Hanany-Witten transitions in 5-brane webs with the mathematical notion of mutation of a Laurent polyonomial/Newton polytope allows us to import the results developed in the mathematical literature. In particular, it has been proven in \cite{2012SIGMA...8..047I,2015arXiv150105334A} that two toric varieties $\mathcal{M}_0$ and $\mathcal{M}_{\infty}$, whose toric diagrams are related by mutation, can be regarded as  deformations of one another. Technically speaking, such deformation is q-Gorenstein (qG), meaning that some power of the canonical bundle is principal (hence trivial as a divisor class). In this language, since the period and the Hilbert series are invariant under qG deformations, their identity under mutation immediately follows. In fact, following \cite{2012SIGMA...8..047I}, one can think of the deformation parameter as a $\mathbb{P}^1$ coordinate, and regard the deformation as a flat fibration over $\mathbb{P}^1$ where the fiber over 0 is $\mathcal{M}_0$ and the fiber over $\infty$ is $\mathcal{M}_{\infty}$. In the particular case of mutations relating $\mathcal{M}_{0,\infty}$ corresponding to toric diagrams (that is, no white dot decorations), it is natural to identify this deformation with the quiver deformation in \cite{Cremonesi:2023psg}.

In practice, it is not trivial to write down the explicit expression for the flat family describing the mutation between two complicated varieties whose brane webs are related by a HW move. Fortunately, in some examples the deformation can be identified with the classical smoothing of singularities. This is in fact the case of our $E_1$ example \cite{2012SIGMA...8..047I}. Focusing on the slice of the dual of the toric fan at height 1, we can simply consider the 2d compact spaces $\mathbb{F}_0$ and $\mathbb{F}_2$ (the construction of the complex cone on top of the 2d compact manifold goes along for the ride in the smoothing to be discussed below). Then, we consider the variety $\mathcal{M}_t$ given by $(y,x,t)\in \mathbb{P}^1\times \mathbb{P}^2\times \mathbb{C}$ with the equation \cite{Huybrechts}
\begin{equation}\label{eq:flat_fibration_E1}
    y_0^2x_0-y_1^2x_1-t\,y_0y_1x_2=0\,.
\end{equation}
For $t\ne 0$ one can solve this equation writing
\begin{equation}
    x_0=(y_1z_0+y_1z_1)^2\,,\qquad x_1=(y_0z_0-y_0z_0)^2\,,\qquad x_2=\frac{4}{t}\,z_0z_1y_0y_1\,.
\end{equation}
This provides a map from $\mathbb{P}^1\times\mathbb{P}^1$ (parametrized by $(z,y)$) into $\mathcal{M}_t$, showing that for any $t\ne 0$, $\mathcal{M}_t$ is isomorphic to $\mathbb{F}_0$. In turn, for $t=0$ the equation boils down to $y_0^2x_0-y_1^2x_1=0$, which describes $\mathbb{F}_2$. Thus we can regard $\mathcal{M}_t$ as a flat fibration of deformations such that the fiber over any arbitrary non-zero $t$ is $\mathbb{F}_0$ with a special fiber at $t=0$ given by $\mathbb{F}_2$. It is interesting to note that the fact the generic fiber is isomorphic to $\mathbb{F}_0$ is consistent with the observation, to be discussed below, that for any non-zero quiver deformation as outlined in \cite{Cremonesi:2023psg}, the theory flows to one describing $\mathbb{F}_0$.

It would be very interesting to make this discussion systematic and to be able to describe the geometry interpolating between any pair of GTPs related by a mutation explicitly. We leave this analysis for future work.

\subsection{Seiberg-Witten curves for GTPs}

As reviewed in Section \ref{sec:HWandMutation}, the vanishing locus of the Laurent polynomial associated to a toric diagram is precisely the SW curve of the 5d SCFT. This gives a physical meaning to the coefficients of the Laurent polynomial as VEVs along the Coulomb branch/mass parameters. Our discussion shows that this conclusion extends to generic GTPs, namely, the SW curve of the 5d theory on the corresponding GTP coincides with the Laurent polynomial associated to the GTP, a suggestion first put forward by \cite{Kim:2014nqa}. The effect of the white dot decoration of the GTP is to reduce the number of independent coefficients. This can be made fully precise for GTPs which arise from mutation of a toric diagram. We will see an explicit example in Section \ref{sec:E0}.

It is natural to ask how to construct the SW curve for a generic GTP with no reference to mutation. Our general discussion instructs us to consider the GTP with no decoration and write down its corresponding Laurent polynomial. For a GTP with $i$ internal points and $b$ boundary points, the number of independent coefficients is $i+b-3$, $i$ of them corresponding to Coulomb branch VEVs and $b-3$ of them corresponding to mass deformations. The GTP decoration freezes some of these coefficients. Through \eqref{eq:dC}, we see that $i-d_C$ Coulomb branch VEVs will be frozen. Moreover, if the GTP has $w$ white points out of the external $b$ points, $w$ masses will be frozen. Alternatively, the number of independent coefficients in the SW for the GTP is $d_c+b-w-3$. 

Which specific mass or Coulomb branch parameters become frozen depends on the locations of the white dots. Recall that that the relation between the complex coefficients in $P(x,y)$ and the masses in the SW curve involves the circle compactification of the brane web and the change of coordinates \eqref{eq:coordinates_SW_curve}. For two points in the toric diagram connected by an edge of length 1, and with associated Laurent polynomial
\begin{align}
    P(x,y) \supset c_1 x^p y^q + c_2 x^{p'}y^{q'} \,,
\end{align}
then the mass resulting from moving the two legs of the web apart from each other is \cite{Aharony:1997bh}
\begin{align}
    M = \frac{R \,T_s}{2\pi} \log\left|\frac{c_1}{c_2}\right|\,,
\end{align}
where $T_s$ is the string tension in type IIB. We see that if we constraint the coefficients to be $c_1=c_2$, the mass is forced to be zero.

Finally, note that in order to write the Laurent polynomial for the GTP before constraining any coefficient, an origin has to be chosen. Such choice must be consistent with the external white dot assignation, even though at this point we do not have a full understanding of the prescription.

\section{BPS Quivers and the Hilbert series}\label{sec:quivers}

So far, our discussion has focused on 5d SCFTs and their geometric realization in M-theory. However, as we reviewed in Section \ref{sec:geoingandwebs}, there is an intimate relation between this and four-dimensional physics: the BPS quiver of the 5d theory realized as M-theory on a CY$_3$ $\mathcal{M}$ coincides with the 4d theory of branes probing $\mathcal{M}$ and, in turn, the moduli space of the 4d theory gives back the CY$_3$ $\mathcal{M}$. It is then natural to wonder whether there is any implication of our work at the level of the BPS quiver. The answer to this question is in the affirmative. As we discussed in Section \ref{sec:invariants}, there is a second invariant under mutations beyond the period, namely the Hilbert series $\text{Hilb}_{\Delta}(t)$ \eqref{eq:def_ehrhart}. It turns out that the BPS quiver knows about this quantity: it coincides with the usual Hilbert Series of its moduli space using the appropriate prescription.

As discussed in Section \ref{sec:invariants}, the Hilbert series of the toric variety can be computed through the Ehrhart series of the dual polytope. Since the Hilbert series is a generating function for the number of holomorphic functions of a given degree in the variety, and since those correspond to gauge-invariant operators (GIOs) in the BPS quiver, it is natural to guess that the partition function for holomorphic GIOs in the BPS quiver should reproduce the Hilbert series of the variety. 

However, this raises two immediate puzzles. First, the Hilbert series, being invariant under mutation, must remain the same for two different-looking CY$_3$s; whereas the GIO partition function would appear to be generally different for distinct CY$_3$s. And second, given a particular toric diagram, one can construct a number of different GTPs depending on the assignation of white dots on the external lines (namely, the choice of external multiplicities in the brane web or, equivalently, the choice of Higgs branch deformations). Indeed, in this way of creating a GTP one finds generically inequivalent 5d theories with different BPS quivers. However, as we have thoroughly discussed, the geometry underlying all these cases is the same, given by regarding the GTP as a standard toric diagram and only freezing some moduli. The first puzzle tells us that one BPS quiver must be associated to several Hilbert series, while the second one says that multiple BPS quivers should be associated to the same Hilbert series of the given geometry.

The solution to both issues turns out to be the same. To compute the Hilbert series of the variety one must assume a particular grading for the holomorphic functions. Likewise, to compute the partition function for GIOs one must assume a particular scaling dimension for each field. It turns out that using the appropriate choice of scaling, the Hilbert series of the variety will coincide with the partition function for GIOs. Moreover, different choices of dimensions in the BPS quiver will allow us to select one GTP or the other within the same undecorated toric diagram.

\subsection{The $E_1$ example} 

To illustrate our discussion, let us consider the $E_1$ example. We have already discussed most of the important ingredients, as it has been our prototype example throughout the paper. As we have seen, we can engineer the $E_1$ theory with M-theory on $C(\mathbb{F}_0)$. Upon mutation, we could equally choose the variety $C(\mathbb{F}_2)$. In fact, we have constructed a fibration of geometries \eqref{eq:flat_fibration_E1} such that the general fiber at any $t\ne 0$ is $\mathbb{F}_0$ which, at $t=0$, becomes $\mathbb{F}_2$. 

\begin{figure}[h]
	\centering
	\includegraphics[height=5cm]{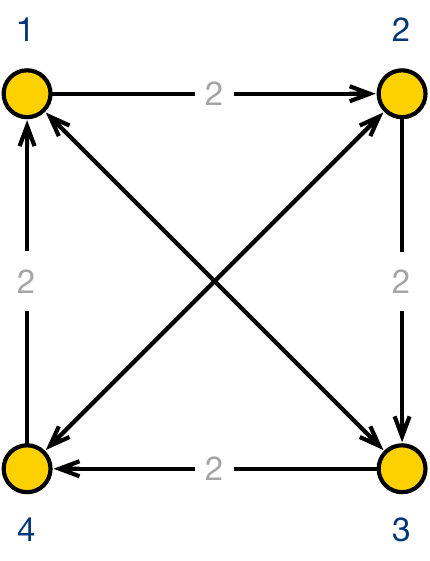}
\caption{Quiver diagram for $C(\mathbb{F}_2)$.}
	\label{quiver_F2}
\end{figure}

Let us now see how this translates into the BPS quiver. The BPS quiver for $C(\mathbb{F}_2)$ is shown in Figure \ref{quiver_F2} \cite{Feng:2004uq}. 

The corresponding superpotential is
\begin{eqnarray}W&=&X_{12}X_{24}X_{41}-X_{23}X_{34}X_{42}+X_{23}Y_{12}X_{31}-X_{12}Y_{23}X_{31} \nonumber \\ && -X_{13}X_{41}Y_{34}+X_{42}Y_{23}Y_{34}+X_{13}X_{34}Y_{41}-X_{24}Y_{12}Y_{41}\,.\end{eqnarray}

We can introduce the following superpotential deformation \cite{Cremonesi:2023psg}

\begin{equation}
\label{defF2}
\delta W=\mu\,(X_{13}X_{31}-X_{42} X_{24})\,,
\end{equation}
which results in the quiver and superpotential for $C(\mathbb{F}_0)$, which were given in Figure \ref{tiling_quiver_F0} and \eqref{W_F0_1}. Note that, as anticipated, for any $\mu\ne 0$, the quiver becomes that for $C(\mathbb{F}_0)$, in accordance with the fact that the generic fiber of the flat fibration smoothing $C(\mathbb{F}_2)$ is $C(\mathbb{F}_0)$.

Let us now turn to the Hilbert series of the BPS quivers. It is straightforward to see that assuming equal dimensions for all fields in the quiver of Figure \ref{tiling_quiver_F0}, we reproduce the Hilbert series in \eqref{eq:HilbE1}, which was computed as the Ehrhart series of the dual polytope. For convenience, we write it again
\begin{equation}
\label{eq:HilbE1_bis}
    {\rm Hilb}_{C(\mathbb{F}_0)}=\frac{1+6t+t^2}{(1-t)^3}\,.
\end{equation}
In particular, this choice of scaling dimensions coincides with the one arising from $a$-maximization \cite{Intriligator:2003jj} had we thought of the BPS quiver as describing a 4d SCFT on D3-branes probing $C(\mathbb{F}_0)$.

In turn, considering the BPS quiver for $C(\mathbb{F}_2)$ in Figure \ref{quiver_F2}, we have to use a different choice of scaling dimensions in order for the Hilbert series to match. If we assume that the fields $\{X_{42},X_{24},X_{13},X_{31}\}$ have dimension 2 and the rest dimension 1 (in the appropriate units), we once again precisely recover \eqref{eq:HilbE1}. Note that with this scaling the superpotential has dimension 4 and the deformation in \eqref{defF2} is (classically) marginal as it has also dimension 4; this is consistent in the sense that when performing a HW move one is not triggering an RG flow.

\begin{figure}[h]
	\centering
	\includegraphics[height=6.5cm]{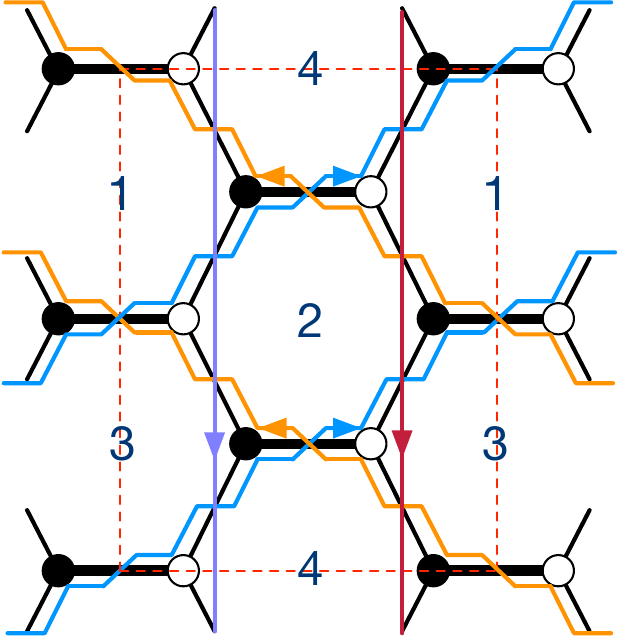}
\caption{Brane tiling for $\C(\mathbb{F}_2)$ with zigzags. Fields with different scaling dimensions are indicated with thick bars.}
	\label{tiling_F2}
\end{figure}

It is useful to identify the fields with special scaling dimensions in the brane tiling. We show these fields with thick bars in Figure \ref{tiling_F2}. We see that the special fields are those not touching the two parallel zigzags, which correspond to the parallel branes in the web.

It is worth stressing that this choice of scaling dimension is not the one arising from $a$-maximization of the 4d theory describing D3-branes probing $C(\mathbb{F}_2)$. In that case, we would have found all dimensions to be equal, leading to
\begin{equation}
\label{HilbF2}
    {\rm Hilb}'_{C(\mathbb{F}_2)}=\frac{1-t+t^2+2t^3+t^4-t^5+t^6}{(1-t)^3\,(1+t+t^2+t^3)^2}\,,
\end{equation}
which is different from the expected result in \eqref{eq:HilbE1_bis}. It is interesting to trace the origin of the disagreement. Since $C(\mathbb{F}_2)$ can be regarded as a $\mathbb{Z}_4$ orbifold of $\mathbb{C}^3$ with action $(1,1,2)$, \eqref{HilbF2} can be computed by a Molien sum of the $\mathbb{C}^3$ Hilbert series. Indeed, one has

\begin{equation}
    {\rm Hilb}'_{C(\mathbb{F}_2)}=\frac{1}{4}\,\sum_{k=0}^3\frac{1}{(1-t\,e^{i\frac{2\pi}{4}k})\,(1-t\,e^{i\frac{2\pi}{4}k})\,(1-t\,e^{i\frac{2\pi}{4}2k})}
\end{equation}

One can check that, instead, \eqref{eq:HilbE1_bis} corresponds to
\begin{equation}
    {\rm Hilb}_{C(\mathbb{F}_2)}=\frac{1}{4}\,\sum_{k=0}^3\frac{1}{(1-t\,e^{i\frac{2\pi}{4}k})\,(1-t\,e^{i\frac{2\pi}{4}k})\,(1-t^2\,e^{i\frac{2\pi}{4}2k})}
\end{equation}
which follows from implementing the same orbifold but on a $\mathbb{C}^3$ where one of the coordinates has twice the dimension of the other tow. This is implemented at the quiver level by the chosen scaling.

\section{Additional examples}\label{sec:examples}

Throughout this paper, we have discussed how mirror symmetry and the mathematical concept of polytope mutations explain how some of the moduli in the extended Coulomb branch of a 5d theory become frozen when the compactification of M-theory is associated to a GTP.  We also investigated the consequences for the corresponding BPS quiver. So far, we have illustrated these points in the context of the $E_1$ theory. In this section, we present several additional examples.

\subsection{The $E_2$ theory} 

\begin{figure}[h]
	\centering
	\includegraphics[height=4.5cm]{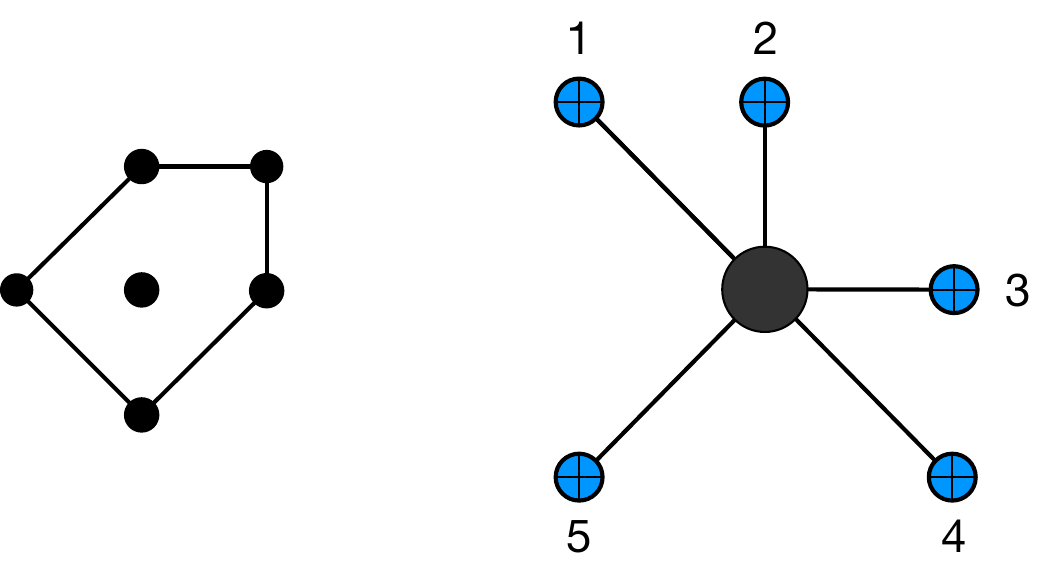}
\caption{Toric diagram and web for $E_2$.}
	\label{toric_web_dP2}
\end{figure}

Let us consider the $E_2$ theory, which can be described by the brane web in Figure \ref{toric_web_dP2}.

\begin{figure}[h]
	\centering
	\includegraphics[height=4.5cm]{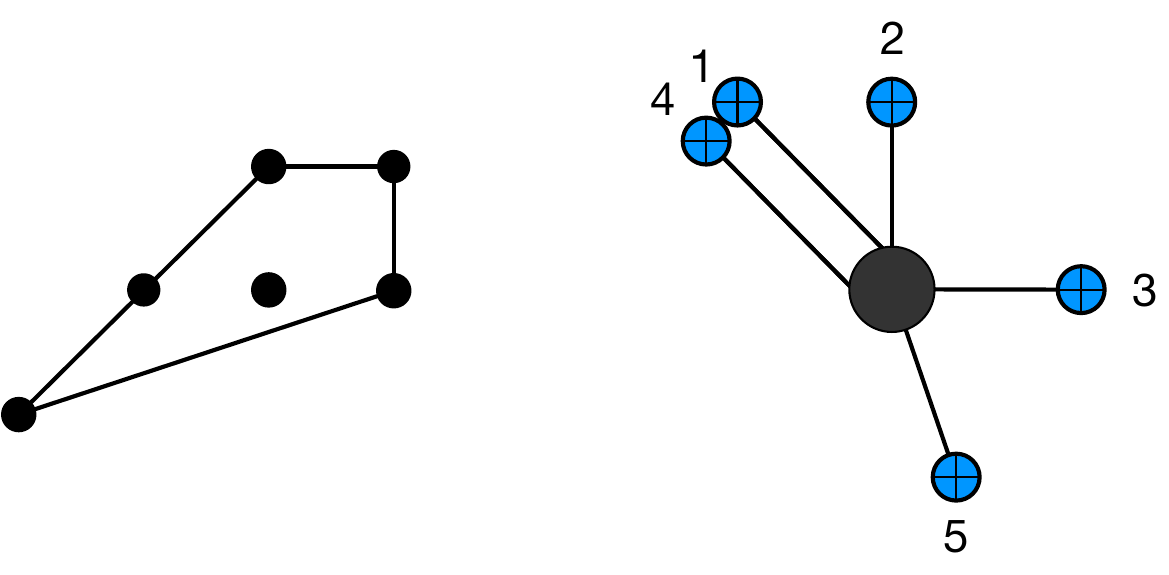}
\caption{Equvalent web for $E_2$ after mutation.}
	\label{toric_web_PdP2}
\end{figure}

As a CY$_3$, Figure \ref{toric_web_dP2} corresponds to $C(dP_2)$. Upon crossing the brane labelled by 4 in Figure \ref{toric_web_dP2}, one obtains Figure \ref{toric_web_PdP2}, which corresponds to $C(PdP_2)$.

One can easily check that the corresponding Laurent polynomials are related by mutation. Indeed, upon SL$(2,\mathbb{Z})$ rotations to appropriately align the toric diagram, the Laurent polynomial for Figure \ref{toric_web_dP2} is
\begin{equation}
    P_{C(dP_2)}(x,y)=\frac{1}{y}\,(c_1 + x)+(c_2+x)+y\,\left(c_3+\frac{1}{x}\right)\,.
\end{equation}
Using the following change of variables $(x,y)\rightarrow (x,y\,(c_1+x))$, one finds
\begin{equation}
    P_{C(PdP_2)}(x,y)=\frac{1}{y}+(c_2+x)+y\,\left(c_1\,\frac{1}{x}+ (c_3c_1+1) + c_3 x\right)\,,
\end{equation}
which is the Laurent polynomial associated to Figure \ref{toric_web_PdP2}. In this case, there are six monomials before and after the mutation and no frozen complex moduli, we have just reshuffled the coefficients. 

Conversely, in terms of the polytope, it is clear that the toric diagrams in Figures \ref{toric_web_dP2} and \ref{toric_web_PdP2} are related by the inversion of a primitive T-cone, which in both cases is of size one, thus resulting in trivial multiplicities after mutation. It is straightforward to compute the Hilbert series associated to the geometries, which, as expected since they are connected by mutation, are equal
\begin{equation}
\label{HSdP2}
{\rm Hilb}_{C(dP_2)}(t)={\rm Hilb}_{C(PdP_2)}(t)=\frac{1+5t^5+t^{10}}{(1-t^5)^3}\,.
\end{equation}

\begin{figure}[h]
	\centering
	\includegraphics[height=5.7cm]{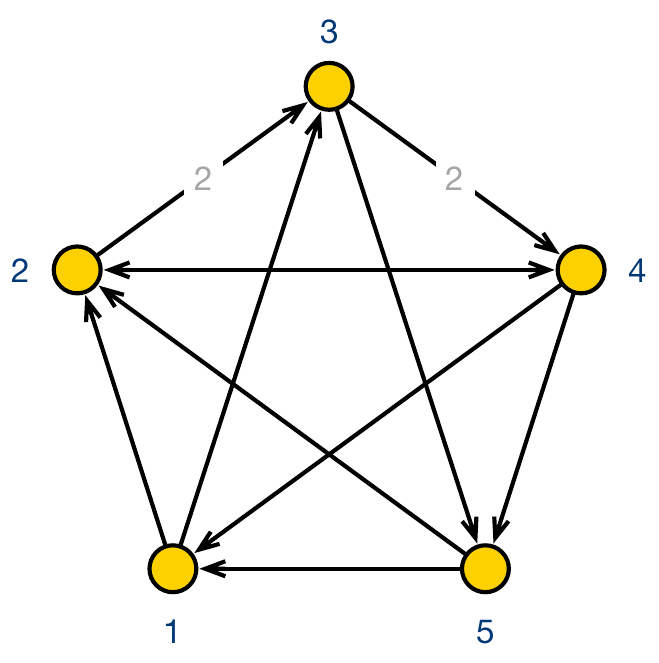}
\caption{Quiver diagram for $PdP_2$.}
	\label{quiver_PdP2}
\end{figure}

Let us now turn to quivers. The BPS quiver for Figure \ref{toric_web_PdP2} is shown in Figure \ref{quiver_PdP2} \cite{Feng:2004uq}.

\begin{figure}[h]
	\centering
	\includegraphics[height=5.7cm]{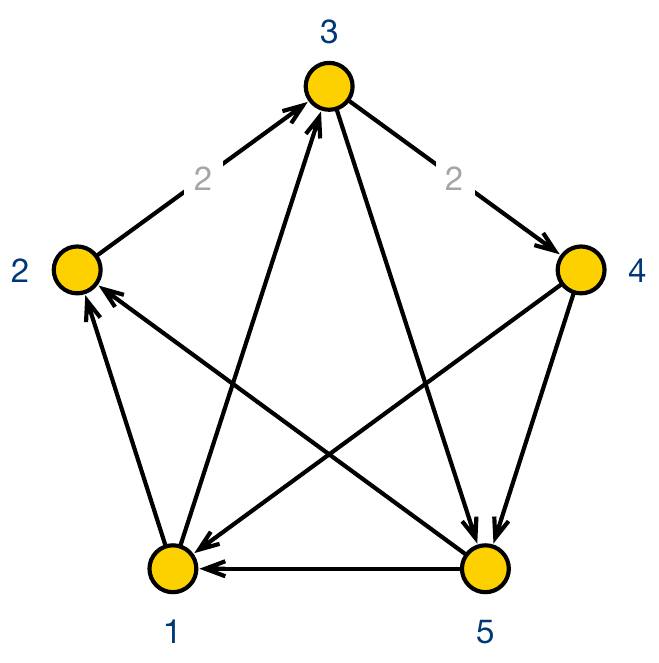}
\caption{Quiver diagram for $PdP_2$.}
	\label{quiver_dP2}
\end{figure}

%
The corresponding superpotential is
\begin{eqnarray}\nonumber W&=&X_{45}X_{51}X_{13}X_{34}+Y_{34}X_{42}Y_{23}+X_{23}X_{35}X_{52}+X_{12}X_{24}X_{41}\\ && -Y_{23}X_{35}X_{51}X_{12}-X_{41}X_{13}Y_{34}-X_{34}X_{42}X_{23}-X_{45}X_{52}X_{24}\,.\end{eqnarray}
Following \cite{Cremonesi:2023psg}, let us introduce the deformation
\begin{equation}
\label{defPdP2}
\delta W=\mu\,(X_{24}\,X_{42}-X_{51}\,X_{13}X_{35})\,,
\end{equation}
and redefine
\begin{equation}
\label{redefinition}
X_{13}\rightarrow \frac{1}{\mu}\,X_{13}-\frac{1}{\mu}\,X_{12}Y_{23}\,,\qquad X_{35}\rightarrow -\frac{1}{\mu}\,X_{35}+\frac{1}{\mu}\,X_{34}X_{45}\,.
\end{equation}
We obtain the quiver in Figure \ref{quiver_dP2} and the superpotential 
\begin{eqnarray}\nonumber W&=&X_{35} X_{51} X_{13} - X_{13} Y_{34} X_{41} - X_{35} X_{52} Y_{23} + Y_{23} X_{34} X_{41} X_{12} \\  && +  X_{52} X_{23} Y_{34} X_{45} - X_{23} X_{34} X_{45} X_{51} X_{12}\,.
\end{eqnarray}
This indeed corresponds to the BPS quiver of the 5d theory specified by the geometry in Figure \ref{toric_web_dP2}.

Let us now turn to the partition function. Starting with the quiver in Figure \ref{quiver_PdP2}, and assuming scaling 3 for $X_{42}$, 2 for $\{X_{41},X_{13},X_{35},X_{52},X_{24}\}$ and 1 for the rest --which is not the assignation which would come from the standard $a$-maximization-- one finds precisely \eqref{HSdP2}. On the other hand, starting with the quiver in Figure \ref{quiver_dP2} and assuming scaling 1 for all fields except $\{X_{41},X_{13},X_{35},X_{52}\}$, which have scaling 2, we again find \eqref{HSdP2}. Once again, we see that the deformation in \eqref{defPdP2}, with the scaling assignations, is classically marginal (and the re-scaling in \eqref{redefinition} homogeneous).

\subsection{The $E_0$ theory}\label{sec:E0}

\begin{figure}[h]
	\centering
	\includegraphics[height=8.5cm]{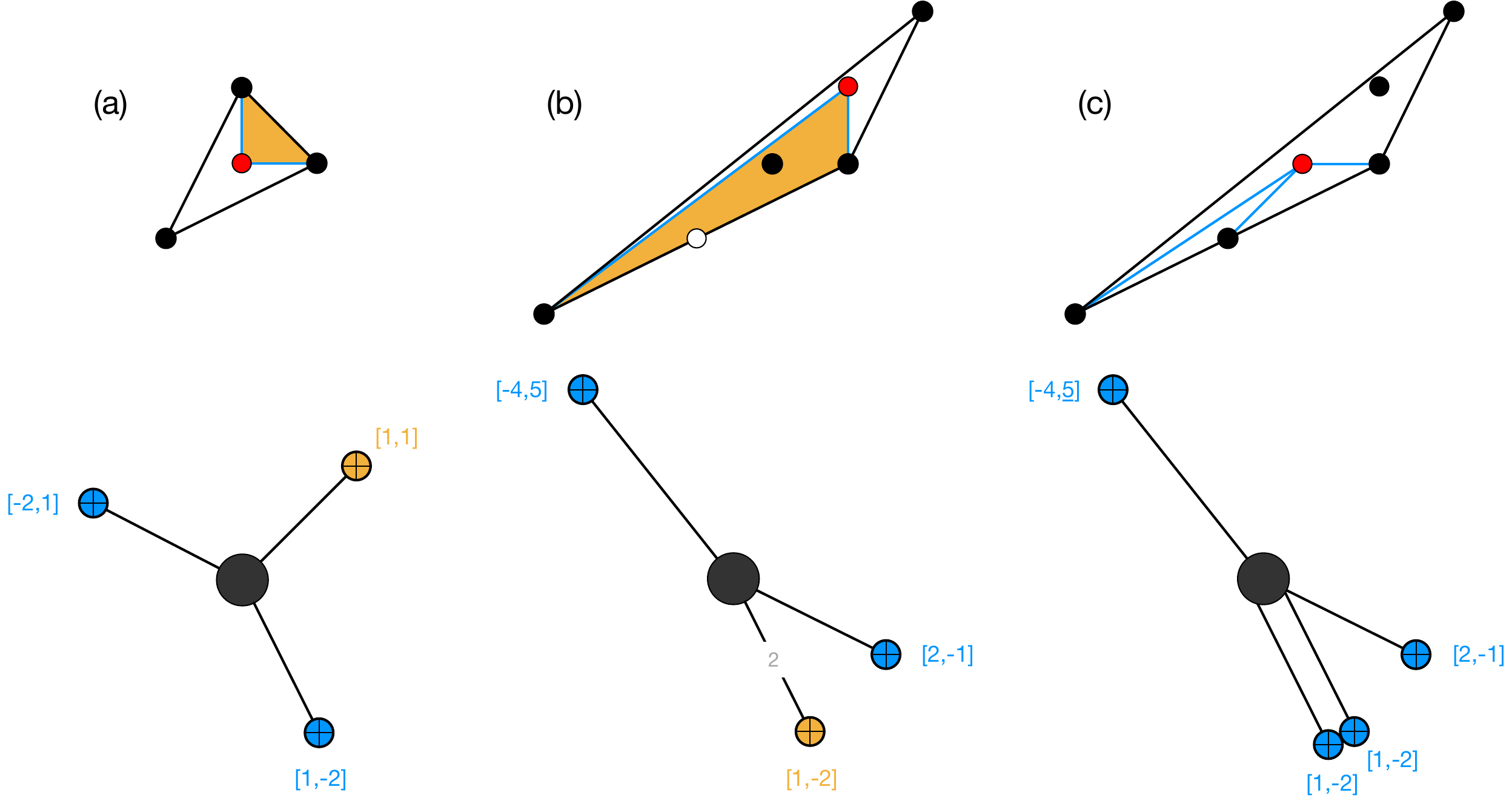}
\caption{(a) Toric diagram and web for the $E_0$ theory. (b) GTP and web for $E_0$ after the mutation, where we see that we have a white dot. (c) Toric diagram and web for SU$(3)_3$ SQCD, which corresponds to the other possible choice of origin of the polytope.}
	\label{toric_web_E0_and_mutations}
\end{figure}

Let us now consider the more involved example of $E_0$, whose web and toric diagram are shown in Figure \ref{toric_web_E0_and_mutations} (a). As a CY$_3$, it corresponds to $C(\mathbb{P}^2)$, also often denoted as $C(dP_0)$.

Following our general discussion, we can choose the leg corresponding to the 7-brane in yellow and cross it to the other side, obtaining Figure \ref{toric_web_E0_and_mutations} (b). This corresponds to a mutation of the polytope where the primitive T-cone in yellow changes side and, more interestingly, size: in this case the resulting brane web has non-trivial multiplicities, as shown Figure \ref{toric_web_E0_and_mutations} (b). At the level of the Laurent polynomial, the starting point is
\begin{equation}
    P_{\text{(a)}}(x,y)=1+c_1x+y+\frac{1}{xy}\,.
\end{equation}

Performing the SL$(2,\mathbb{Z})$ transformation $(x,y)\rightarrow (\frac{x}{y},\frac{1}{y})$ to align the side to be mutated with the $x$-axis, and performing the mutation $(x,y)\rightarrow (x,y\,(1+c_1 x))$, the result is
\begin{equation}
\label{SWE0GTP}
    P_{\text{(b)}}(x,y)=\frac{1}{y}+1+\big(2c_1+\frac{1}{x}+c_1^2x)\,y^2\,.
\end{equation}
We recognize the Laurent polynomial for the polytope in Figure \ref{toric_web_E0_and_mutations} (b), with three of its moduli frozen to specific values in terms of $c_1$ (one of them equal to zero). Moreover, one can check that indeed the Hilbert series --computed as the Ehrhart polynomial of the corresponding $\Delta^{\circ}$ for both Figures \ref{toric_web_E0_and_mutations} (a) and (b) with the given origin-- agrees and reads
\begin{equation}
\label{ZC3modZ3}
{\rm Hilb}_{\text{(a)}}(t)={\rm Hilb}_{\text{(b)}}(t)=\frac{1+7t+t^2}{(1-t)^3}\,,
\end{equation}
It is also straightforward to check the invariance of the period, i.e. $\pi_{P_{\text{(a)}}}(t)=\pi_{P_{\text{(b)}}}(t)$.

This case actually corresponds to a GTP. Imagine now changing the origin to the other internal point as shown in Figure \ref{toric_web_E0_and_mutations} (c). Then, the primitive cones subtended are of length 1, which means that now the external multiplicities are all 1 as shown in the web in Figure \ref{toric_web_E0_and_mutations} (c) (which, as a 5d theory, corresponds to $SU(3)_3$ SQCD). Computing the Hilbert series for this case through the Ehrhart series, one finds
\begin{equation}
    \label{SU(3)_3}
    {\rm Hilb}_{\text{(c)}}(t)=\frac{1+3t+10t^2+3t^3+t^4}{(1-t)^3\,(1+t)^2} \,.
\end{equation}
Clearly ${\rm Hilb}_{\text{(a)}}(t)\ne {\rm Hilb}_{\text{(c)}}(t)$, in accordance with the fact that the webs (a) and (c) in Figure \ref{toric_web_E0_and_mutations} are not related by mutation.

\begin{figure}[h]
	\centering
	\includegraphics[height=5cm]{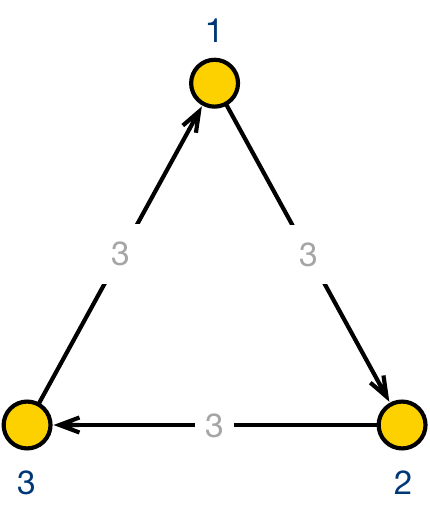}
\caption{Quiver diagram for $C(dP_0)$.}
	\label{quiver_dP0}
\end{figure}

Let us now turn to quivers. The original quiver for the Figure \ref{toric_web_E0_and_mutations} (a) is that for D3-branes probing $C(dP_0)$, and is shown in Figure \ref{quiver_dP0} \cite{Feng:2001xr,Feng:2002zw}.

The superpotential for this theory is
\begin{equation}
W=\epsilon_{ijk}X_{12}^iX_{23}^jX_{31}^k\,.
\end{equation}

It is straightforward to compute the partition function for this theory assuming equal scaling for all fields, finding precisely \eqref{ZC3modZ3}.

\begin{figure}[h]
	\centering
	\includegraphics[height=6.5cm]{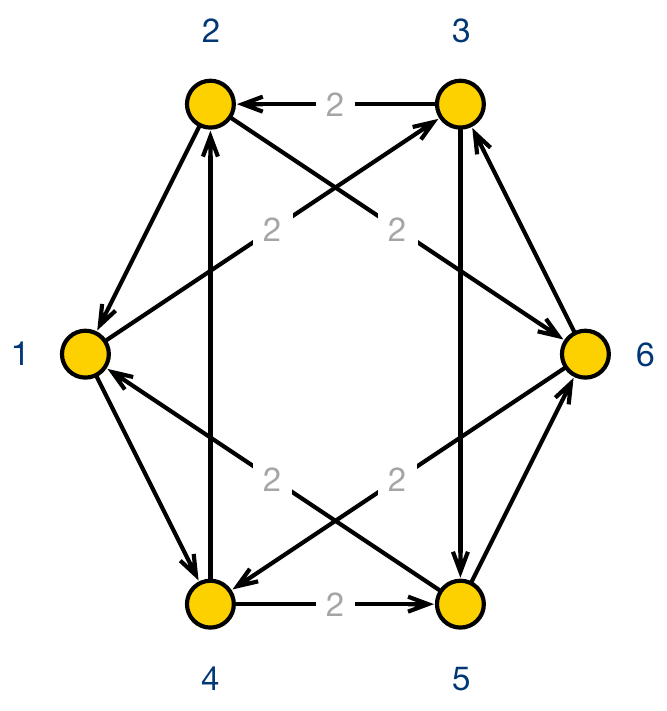}
\caption{Quiver diagram for the $\mathbb{C}^3/\mathbb{Z}_2\times \mathbb{Z}_3$ orbifold with weights $(1,-1,0)$ and $(1,1,1)$.}
	\label{quiver_Z2_Z3}
\end{figure}

Let us now turn to Figure \ref{toric_web_E0_and_mutations} (b). As thoroughly argued, the corresponding geometry is the one associated to the GTP regarding it as an ordinary toric diagram, neglecting all white dot decorations. In this particular case, it is the toric diagram of $\mathbb{C}^3/\mathbb{Z}_2\times \mathbb{Z}_3$ where the orbifold has weights $(1,-1,0)$ and $(1,1,1)$. Its quiver is shown in Figure \ref{quiver_Z2_Z3} \cite{Franco:2017jeo}.

The superpotential for this theory is
\begin{eqnarray}\nonumber W&=&X_{56}X_{64}X_{45}+X_{63}X_{32}X_{26}+Y_{13}X_{35}Y_{51}+X_{14}Y_{45}X_{51}\\  && +X_{21}X_{13}Y_{32}+Y_{26}Y_{64}X_{42}  - X_{64}X_{42}X_{26}-Y_{13}X_{32}X_{21}\\ \nonumber && -Y_{51}X_{14}X_{45}-Y_{64}Y_{45}X_{56}-X_{51}X_{13}X_{35}-Y_{32}Y_{26}X_{63}\,.
\end{eqnarray}

\begin{figure}[h]
	\centering
	\includegraphics[height=8.5cm]{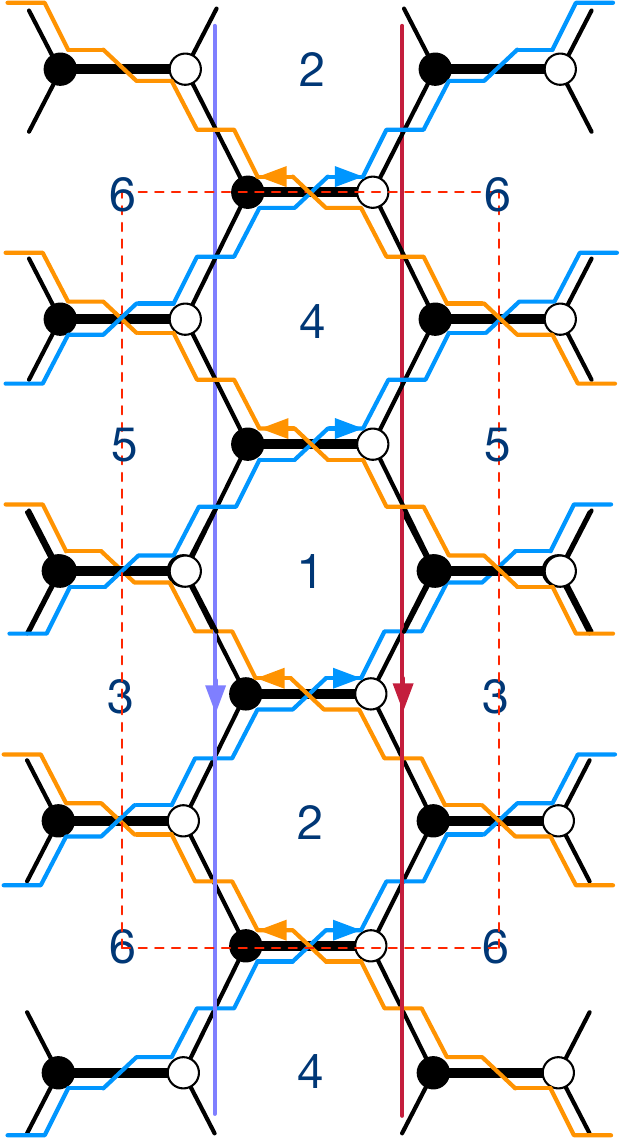}
\caption{Brane tiling for $\mathbb{C}^3/\mathbb{Z}_2\times \mathbb{Z}_3$ with zigzags. Fields with different scaling dimensions are indicated with thick bars.}
	\label{tiling_Z2_Z3}
\end{figure}

One can see that assigning scaling 4 to $\{X_{21}, X_{14}, X_{42}, X_{35}, X_{56}, X_{63}\}$ and 1 to the rest of the fields, the partition function of the quiver is given by \eqref{ZC3modZ3}. In this case, as one side of the mutation involves a true GTP, the deformation relating the quivers in Figures \ref{quiver_dP0} and \ref{quiver_Z2_Z3} is not known. Once again, it is interesting to identify the fields with special scaling dimensions in the brane tiling \cite{Franco:2017jeo}. We show them with thick bars in Figure \ref{tiling_Z2_Z3}. Once again, we see that the special fields are those that do not intersect the two parallel zigzags, which correspond to the parallel branes in the web.

According to our general discussion, the quiver in Figure \ref{quiver_Z2_Z3} must also correspond to Figure  \ref{toric_web_E0_and_mutations} (c). The difference must lie in the scaling dimensions. Indeed, one can now check that, upon setting all dimensions equal, the partition function for the quiver is precisely \eqref{SU(3)_3}.

We can trace the geometric origin og the chosen dimensions. The polytope in Figure \ref{toric_web_E0_and_mutations} (b) and (c) corresponds to the toric diagram of $\mathbb{C}^3/\mathbb{Z}_2\times \mathbb{Z}_3$ where the orbifold has weights $(1,-1,0)$ and $(1,1,1)$. Hence, assuming equal scalings for all the $\mathbb{C}^3$ coordinates one finds
\begin{equation}
    {\rm Hilb}_{\text{(c)}}(t)=\frac{1}{2}\sum_{u^2=1}\frac{1}{3}\sum_{v^3=1}\frac{1}{(1-t\,uv)\,(1-t\,\frac{v}{u})\,(1-t\,v)}\,.
\end{equation}
In turn,
\begin{equation}
    {\rm Hilb}_{\text{(b)}}(t)=\frac{1}{2}\sum_{u^2=1}\frac{1}{3}\sum_{v^3=1}\frac{1}{(1-t\,uv)\,(1-t\,\frac{v}{u})\,(1-t^4\,v)}\,.
\end{equation}

\subsection{The $\tilde{E}_1$ theory} 

\begin{figure}[h]
	\centering
	\includegraphics[height=4.5cm]{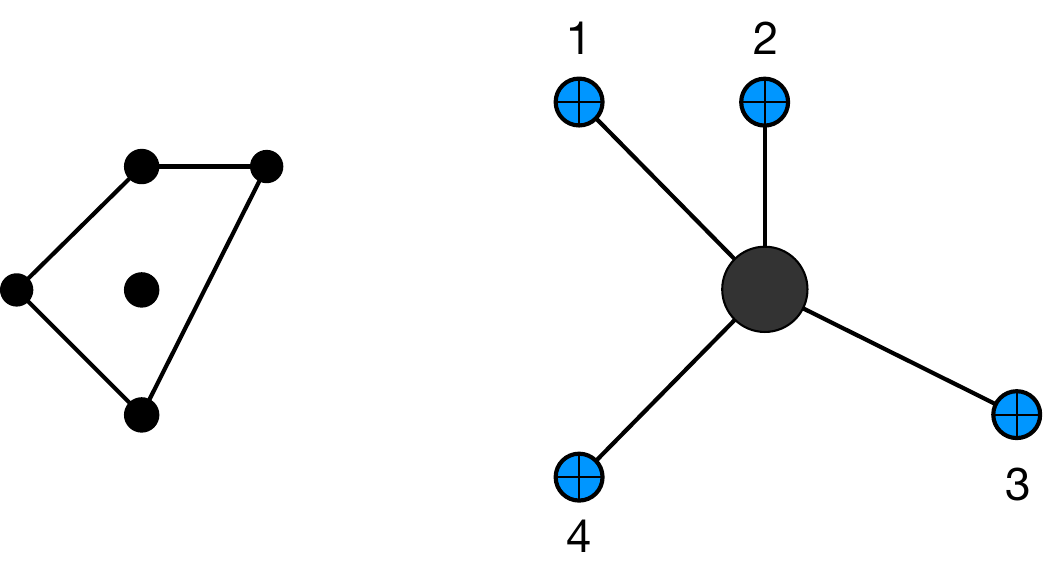}
\caption{Web for the $\tilde{E}_1$ theory.}
	\label{toric_web_dP1}
\end{figure}

Let us now consider the $\tilde{E}_1$ theory, which can be encoded in the toric diagram and web in Figure \ref{toric_web_dP1}. The toric diagram corresponds to the cone over $dP_1$.

Computing the Hilbert series of the geometry as the Ehrhart series of the dual polytope one finds,
\begin{equation}
\label{HilbdP1}
{\rm Hilb}_{C(dP_1)}=\frac{1+6t^4+t^8}{(1-t^4)^3}\,,
\end{equation}

\begin{figure}[h]
	\centering
	\includegraphics[height=5cm]{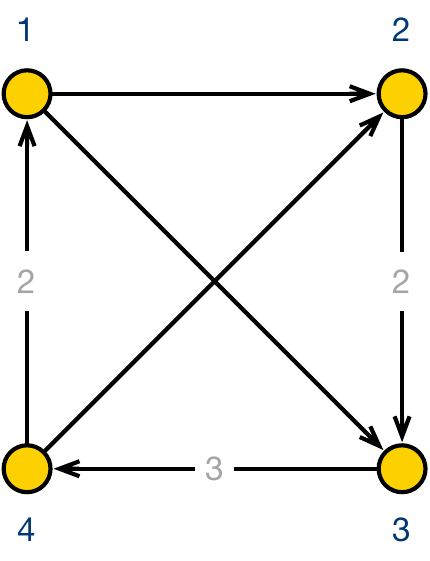}
\caption{Quiver diagram for $dP_1$.}
	\label{quiver_dP1}
\end{figure}

Figure \ref{quiver_dP1} shows the BPS quiver for $dP_1$ \cite{Feng:2001xr,Feng:2002zw}.

The corresponding superpotential is
\begin{eqnarray}\nonumber W&=&X_{13}Z_{34}Y_{41}-X_{12}X_{23}X_{34}Y_{41}+X_{12}X_{34}X_{41}Y_{23}\\ \nonumber && -Z_{34}X_{42}Y_{23}-X_{13}X_{41}Y_{34}+X_{23}X_{42}Y_{34}\,.
\end{eqnarray}
 
One can check that, assigning scaling 2 to $\{X_{42},X_{13}\}$ and 1 to the rest of the fields, one recovers \eqref{HilbdP1}.

\section{Discussion}\label{sec:discussion}

We have made progress towards the understanding of the geometrical engineering of generic 5d SCFTs and its relation to brane webs, thus enlarging the class of theories for which both descriptions are available. The key observation is that HW transitions in brane webs correspond to simple coordinate transformations in the geometry of the mirror, which in turn allows us to make contact with the mutations introduced in the mathematical literature. This new connection has very interesting consequences for the Physics of 5d SCFTs (and may as well have implications for pure Mathematics through new insights from Physics). Let us briefly summarize our findings. 

There exists a mathematical notion of mutation, which can be expressed either at the level of a polytope or its associated Laurent polynomial. This mutation transforms the polytope/polynomial according to certain mutation data, which encodes the choice of HW transition of the corresponding brane web. Regarding the polytope as the toric diagram of a toric variety, this transformation can be viewed as a deformation connecting two toric varieties. Moreover, it is such that the periods and the Hilbert series remain invariant. An important observation is that the choice of origin is crucial when implementing the mutation. From a polytope perspective, the mutation amounts to selecting a primitive T-cone and reversing it. In terms of the Laurent polynomial, the choice of origin requires tuning a number of coefficients --mirroring the primitive T-cone-- in the polynomial to implement the mutation. 

Our observation is that polytope mutations correspond to HW transitions in webs of 5-branes ending on 7-branes. As usual, the graph dual to a brane web is a GTP.  Then, HW transitions correspond to mutations of the GTP upon identifying the 7-brane to cross, along with the attached 5-branes, with the primitive T-cone. In this identification, the length of the base of the cone is the multiplicity of the leg, while the vector normal to the base of the cone is the $[p,q]$ charge of the 7-brane to cross. A consequence of this correspondence is that the GTP should be regarded as a standard toric diagram, and that the decoration with white dots encodes the choice of origin through the selection of primitive T-cones. For trivial external multiplicities, it was well-known that the GTP is the toric diagram of the toric CY$_3$ that geometrically engineers the 5d SCFT in M-theory. In this paper, we have seen that this extends to arbitrary GTPs. The geometry that engineers these theories in M-theory is the one defined by the GTP interpreted as a toric diagram, i.e. forgetting about its white dot decoration. The effect of white dots is to freeze some of its moduli.

The correspondence between mutations and HW transitions imply that the period and the Hilbert series are invariant. We conjecture that their inclusion results in a complete set of classifiers of brane webs, going beyond the proposal in \cite{DeWolfe:1998eu}. Indeed, we used them to resolve a puzzle raised in \cite{Arias-Tamargo:2022qgb}. Motivated by these findings, it would be very interesting to perform a classification attempt similar to the one in  \cite{Arias-Tamargo:2022qgb}, but based on the invariance of the period and the Hilbert series.

We have seen that the Hilbert series can be computed from the BPS quiver, which is formally identical to the fractional brane quiver for D3-branes probing the associated toric geometry. Remarkably, this approach provides further quantitative evidence supporting the conclusion that the geometry associated with a GTP is equivalent to considering the GTP as a toric diagram. Notably, the choice of origin, and thus the selection of the white dot decoration, is encoded in the scaling of the fields when computing the GIO partition function. This aspect was explicitly demonstrated in Section \ref{sec:E0}, where the fractional brane quiver for the toric diagram on the right of Figure \ref{toric_web_E0_and_mutations} reproduces, depending on the chosen field scaling, either the Hilbert series for the toric diagram or that for the GTP in the middle of Figure \ref{toric_web_E0_and_mutations}, corresponding respectively to the $SU(3)_3$ 5d SCFT or the $E_0$ theory.

As we have seen, white dots encode constraints on the possible resolutions of the geometry, i.e. Coulomb branch VEVs and masses. This becomes evident when considering the mutation of the Laurent polynomial. Indeed, starting with a web with trivial external multiplicities, mutation typically generates a GTP. However, it is clear that the number of coefficients in the Laurent polynomial remains fixed under mutation. Consequently, the resulting Laurent polynomial possesses fewer independent coefficients than a priori allowed. 

Since the SW curve is the zero locus of the Laurent polynomial \cite{Aharony:1997bh}, it follows that the effect of white dots is to freeze some of its deformations, thus recovering the results in \cite{Kim:2014nqa} from first principles.

\subsection{Open questions}

Our work suggests various interesting directions for future research. Below, we summarize some of them.

While our work shed light on the importance of the choice of origin for toric diagrams, its complete implications remain to be understood. For example, a toric diagram with $n_i$ collinear edges on the $i$-th side can result in numerous distinct GTPs, each corresponding to different possible (supersymmetric) white dot decorations. Currently, it is unclear whether all these possibilities can be solely encoded by the choice of origin, and if so, how. A related question concerns the status of non-primitive vectors. In the mathematical literature, position vectors of polytope corners are typically required to be primitive, whereas Physics suggests the necessity of allowing non-primitive corners. For instance, the toric diagram dual to the the web for $SU(N)_k$ SQCD has $N-1$ internal collinear points. This implies that for $N\geq 2$, no choice of origin results in primitive vectors for the corners. In this case, it is not clear what the different choices of origin mean. Note that for each choice, the Hilbert series computed through the Ehrhart series of the dual polytope yields a different result. Additionally, in this case, the choices cannot correspond to different multiplicities.

Another surprising finding of this paper is that the choice of origin is encoded in the BPS quiver through the scaling of fields. Although we have presented evidence for this, understanding the rationale behind the scaling choice from first principles remains unclear. In other contexts, such as SCFTs in 4d, 3d, and 2d, the scaling is determined by an extremization principle arising from an RG flow (e.g., $a$-maximization for 4d SCFTs). While in the case at hand we are dealing with a BPS quiver, which should be understood as Matrix Quantum Mechanics, one might also anticipate a dynamical assignment of dimensions. It would be interesting, albeit challenging, to investigate whether some prescription can produce multiple GTPs (which arise from white dot decorations of a single underlying toric diagram) from a unique quiver.

In all examples that we have examined where there is a mutation relating two standard toric diagrams (i.e., without white dots), the corresponding BPS quiver theories are connected by a superpotential deformation of the type studied in \cite{Cremonesi:2023psg}, which, with our scalings, is classically marginal.\footnote{In our convention, this means that $\delta W$ has the same scaling as the original superpotential.} It is natural to conjecture that it is actually marginally relevant, so that it triggers a flow to the IR geometry. This resonates with Ilten’s geometric perspective on the deformation, which is described in terms of a flat fibration \cite{2012SIGMA...8..047I}). For instance, in the $E_1$ example, for any non-zero deformation parameter, the generic fiber is $\mathbb{F}_0$, while only at the origin do we find $\mathbb{F}_2$. It would be interesting how to extend this picture to the non-toric case.\footnote{For instance, in the $E_0$ case, it would be natural to consider a deformation of the form $\delta W= \mu_1\, X_{32}X_{26}X_{64} X_{45} X_{51} X_{13}+\cdots$ of Figure \ref{quiver_Z2_Z3}, which is marginal with the scalings in the text. It is unclear how this would lead to Figure \ref{quiver_dP0}.}

Additionally, in contrast with the HW transition, the deformation of the BPS quiver naively seems to be an irreversible process, in harmony with the RG flow picture.\footnote{Allowing for massive fields to be integrated in, etc, it is in principle possible to imagine going in both directions.} The reversibility of the process is naturally implemented in the twin perspective as advocated in \cite{Franco:2023flw,Franco:2023mkw}, where the HW transition corresponds to (formal) Seiberg duality in a (not necessarily toric) node of the twin quiver. This is also related to the natural question of the fate of the various (toric) phases of the BPS quiver, which from the twin point of view has been discussed in \cite{Franco:2023mkw}. It would be interesting to flesh out the mapping between all these different viewpoints.

Finally, it would be interesting to explore the geometric side of the mutation in further detail, explicitly constructing the flat family of geometries fibered over a $\mathbb{P}^1$. It is natural to ask whether it is possible to write a metric for the generic element of such interpolating fibration, and wether extra fluxes are present. In addition, it would be nice to establish the dictionary between the deformation of the BPS quiver and the geometric deformation, at least in simple examples.

\section*{Acknowledgements}

We would like to thank Antoine Bourget, Tom Coates, Amihay Hanany and Mario De Marco for useful discussions. G.A.T. is supported by the STFC Consolidated Grants ST/T000791/1 and ST/X000575/1. S.F. is supported by the U.S. National Science Foundation grants PHY-2112729 and DMS-1854179.  D.R.G is supported in part by the Spanish national grant MCIU-22-PID2021-123021NB-I00.

\printbibliography

\end{document}